\documentclass[print]{elsarticle}
\pdfoutput=1
\usepackage{lineno,hyperref}
\usepackage{graphicx}
\usepackage[linesnumbered,ruled,lined]{algorithm2e}
\usepackage{float}
\usepackage{tikz}
\usepackage{epstopdf}
\usepackage{amssymb}
\usepackage{amsmath}
\usepackage{subcaption}
\usepackage{footnote}
\usepackage{array}
\usepackage{cases}
\usepackage{multirow}
\usepackage{color}
\journal{CMAME}
\definecolor{chiColor}{rgb}{255,0,0}

\begin{document}
%
%
\begin{frontmatter}
		\title{Level-set based pre-processing algorithm for particle-based methods}
		\author[myfirstaddress]{Yongchuan Yu }
		\ead{yongchuan.yu@tum.de}
		\author[mysecondaryaddress,mythirdaddress]{Yujie Zhu }
		\ead{yujie.zhu@tum.de}
		\author[mysecondaryaddress]{Chi Zhang }
		\ead{c.zhang@tum.de}
		\author[myfirstaddress]{Oskar J. Haidn}
		\ead{oskar.haidn@tum.de}
		\author[mysecondaryaddress]{Xiangyu Hu \corref{mycorrespondingauthor}}
		\ead{xiangyu.hu@tum.de}
		\address[myfirstaddress]{Chair of Space Propulsion and Mobility, 
			Technical University of Munich, 85521 Ottobrunn, Germany}
		\address[mysecondaryaddress]{Chair of Aerodynamics and Fluid Mechanics, 
			Technical University of Munich, 85748 Garching, Germany}
		\address[mythirdaddress]{Xi'an Research Institute of High-Tech, 
		710025 Xi'an, China}
		\cortext[mycorrespondingauthor]{Corresponding author. }
\begin{abstract}
Obtaining high quality particle distribution representing clean geometry
in pre-processing is essential 
for the simulation accuracy of the particle-based methods. 
In this paper,
several level-set based techniques for cleaning up `dirty' geometries automatically 
and generating homogeneous particle distributions are presented.
First, a non-resolved structure identifying method based on level-set field is employed
to detect the tiny fragments which make the geometry `dirty' under a given resolutions.
Second, a re-distance algorithm is proposed to remove the tiny fragments 
and reconstruct clean and smooth geometries.
Third, a `static confinement' boundary condition 
is developed in the particle relaxation process.
By complementing the kernel support for the particles near the geometric surface,
the boundary condition achieves better body-fitted particle distribution 
on the narrow region with high curvature.
Several numerical examples include a 2D airfoil 30P30N, 
3D SPHinXsys symbol, a skyscraper with a flagpole and an inferior vena cava
demonstrate that the present method not only cleans up 
the `dirty' geometries efficiently, 
but also provides better body-fitted 
homogeneous particle distribution for complex geometry.
\end{abstract}

\begin{keyword}
Particle methods \sep `dirty' geometry cleaning\sep level-set \sep static confinement \sep kernel support completing
\end{keyword}
\end{frontmatter}
%
%
\section{Introduction}\label{sec:introduction}
As a truly Lagrangian, mesh-free method, 
smoothed particle hydrodynamics (SPH) has attracted 
tremendous attention due to its very nature of tracking moving characteristics such as free surfaces,
moving and deformable material interfaces. 
Typical applications include geophysical flow
\cite{Industrial--xenakis2017,Industrial--cleary2020,Industrial--cleary2012}, 
bio-mechanics \cite{Industrial--zhang2021,Industrial--harrison2016,Industrial--tanaka2005, bio--zhang2022artificial}
and other industrial application 
\cite{Industrial--leroy2014,Industrial--groenenboom2019,Industrial--shadloo2016}.
Specifically, the SPH method has been successfully implemented 
in the modeling of bird impact \cite{Industrial--lavoie2008,distribute--siemann2019},  
aircraft ditching \cite{Industrial--Ortiz2004}, 
wave energy conversion process \cite{Industrial--zhang2020}, 
ocean and coastal engineering \cite{ocean--luo2021particle, ocean--cai2022sph,Industrial--pan2016},
high-velocity impact welding \cite{Industrial--NASSIRI2016}, 
and slurry$\And$media motion within stirred media detritor (SMD) \cite{Industrial--NDIMANDE2019}, 
to name a few. 

To eliminate the bottleneck for widespread industrial applications of SPH method, 
one critical challenging task in pre-processing 
is to efficiently generate body-fitted and isotropic particle distribution for arbitrarily complex geometry. 
In many applications of the SPH method, 
the lattice-based 
and volumetric-mesh converted particle distributions \cite{Pre-process-distribute--Dominguez2011,distribute--wang2022centrifugal} are still 
the most popular approaches \cite{Industrial--shadloo2016}. 
While the former is not body-fitted for complex geometry, 
the latter is difficult to be isotropic.
When SPH particles are used to model a fluid in the Lagrangian framework,  
it seems that the initial particle distribution 
does not need to be body-fitted or isotropic as,
after the simulation starts, the particles always 
leave their initial position and form isotropic distribution 
due to the self-adjust mechanism. Even this, since the particles are not
initially set in the `equilibrium' position, the numerical noise introduced 
by particle resettlement during the early stages of the flow evolution 
will strongly affect the fluid evolution \cite{distribute--colagrossi2012}.
Furthermore, as the development 
of more challenging SPH algorithms and complex applications in the field of bio-medicine\cite{Industrial--zhang2021}, 
structural mechanics\cite{Industrial--peng2019}, fluid-structure interaction \cite{FSI--han2018,FSI--liu2019,FSI-MR--zhang2021}, 
and SPH method in Eulerian framework \cite{Eulerian--fourtakas2018, Eulerian--nasar2019}, 
in which the particle configuration is fixed through the simulation, 
generating body-fitted and isotropic particle distribution 
for general complex geometries are becoming more significant.

In order to achieve this goal, 
several different approaches have been proposed.
Particle packing algorithm \cite{distribute--colagrossi2012} provide a `equilibrium' 
initial distribution of fluid particles with simple geometries 
where solid particles are prescribed for the surface. 
The weighted Voronoi tessellation (WVT) algorithm \cite{distribute--2015optimal-initial} 
iterates the particle distance under a repulsive force 
to achieve a quasi-isotropic uniform or non-uniform  
particle distribution for arbitrary geometry. This method needs to exert 
anti-symmetric forces from the ghost particles to prevent the penetration. 
The Extended WVT algorithm 
\cite{distribute--PHL.Groenenboom2014Extended-WVT, distribute--siemann2014modeling, distribute--siemann2019} 
represents the geometry surface by shell mesh elements in order to treat more complex boundaries. 
By introducing level-set to describe the geometry while solving the target feature-size
function, Fu et al. \cite{MR--fu2019} proposed a fluid relaxation method, which is able to generate 
isotropic and body-fitted particle distribution for arbitrary 2D geometry . 
Ji et al. \cite{distribute-MR--ji2020, MR-distribute--JI2021} 
extended the fluid relaxation method with a feature boundary correction term to accelerate the particle generation process 
and to address the issue of incomplete kernel support near the surface.
By replacing the governing equation with an original momentum equation in particle discrete form, 
Zhu et al.\cite{distribute--yujiezhu2021} further simplified the relaxation method to a 
physics-driven relaxation process with a simple level-set based bounding method. 

While the relaxation-based methods are able to generate body-fitted 
and isotropic particle distributions,  
when the local length scale or curvature of the geometry
does not variate too much (except singularities), 
they face a critical issue,
similar to the mesh-based method \cite{mesh-generate--chawner2016},
when the local length scales of the body surface span a large range. 
One typical case is the problem of `dirty' geometry, 
which frequently occurs in industrial simulations 
and brings lots of extra manual labor, 
but has been rarely discussed in the literature.  
`Dirty' geometries involve small structures or 
small flow paths with the characterized size  
which can not be well resolved in the simulations \cite{clean--keiji2021}, 
and often induce low-quality mesh or particle distribution 
and eventually to numerical instability.
Besides erasing these small structures manually,    
another straightforward approach is capturing such small features directly, 
which may increase the total number of grid points or particles dramatically 
and leads to extremely small time-step sizes \cite{clean--keiji2021}. 
One alternative way to circumvent this issue for the mesh-based method is introduced 
an effective smearing technique by which 
the influence of small structures is smeared out 
with the immersed boundary method (IBM) \cite{clean--keiji2021}. 
While this technique shows good performance in many cases, 
the overlap meshes and heterogeneous coupling used in IBM 
are not always desirable,
especially when conservation properties are required.  
To the best knowledge of the authors, 
the proper approach for handling the `dirty' geometry problem is 
yet to be proposed for particle methods.
 
Another issue of the particle relaxation method is 
the boundary condition for body-fitting. 
As the particle relaxation takes place inside the geometry, 
and there are no particles outside the surface to achieve 
force balance or full kernel support,
a boundary condition should be imposed for the particles near the surface. 
A typical approach for this is the ghost particle method, 
in which ghost particles are used to fill the outside space near 
the surface \cite{boundary--2017, boundary--schechter2012, boundary--vela2019, MR--fu2019}. 
While introducing ghost particles for simple surfaces is straightforward, 
it can be quite a challenge for complex geometry. 
In addition, more particles also lead to additional 
computational and memory cost, especially for three-dimensional problems. 
Ji et al. \cite{MR-distribute--JI2021} 
exploited a `feature boundary correction' term to mimic the 
fully kernel support for the boundary particles near the surface. 
This method still requires a single layer of particles generated
on the surface,
which is not trivial for complex three-dimensional 
geometries, and relaxing together with the interior particles.    
Zhu et al. \cite{distribute--yujiezhu2021} 
introduced a simple and fast particle bounding method 
without using ghost or surface particles. 
The bounding method directly constrains the boundary particles
according to their distances toward the surface 
probed from the background level-set field.    
While being quite effective when the surface curvature is moderate, 
it does not converge, 
typically presented by particles with persistent and fast motion, 
to the balanced particle distribution near sharp features of the surface. 

In this paper, a level-set based pre-processing techniques for 
particle-based applications are proposed to solve the above-mentioned problems. 
First, a non-resolved structure identifying method based on 
level-set field is employed to find out those tiny fragments which make
the geometry becomes `dirty' under a given resolution.
Second, a re-distance method is used to reconstruct 
the level-set field by removing these `dirty' geometries.
The particle generation process for the 30P30N airfoil is employed 
to show the instability caused by `dirty' geometry in particle relaxation.
This example also demonstrates that our method generates the 
isotropic body-fitted particle distribution for arbitrary complex 
geometries by identifying and cleaning-up the non-resolvable 
small structures.
At last, a new level-set based method named `static confinement' is 
developed to complete the kernel function on the 
geometry surface during the physics-driven relaxation process. 
With the `static confinement' boundary condition, 
the geometries with large curvature corners can be accurately captured.
      
The remainder of this paper is organized as follows.
In Section \ref{sec:previous-work}, 
we briefly summarize the previous particle-relaxation method 
for body-fitted particle generation and the narrow-band level-set technique.  
In Section \ref{sec:clean}, 
the algorithm for identifying and cleaning-up 
the non-resolvable small structures are presented.
The method of `static confinement'  
is developed in Section \ref{sec:static confinement} 
to complete the SPH kernel support during the particle relaxation process. 
Several typical applications are shown in Section \ref{sec:numerical examples} to validate 
the importance of the `dirty' geometry cleaning-up process and the 
self-cleaning ability of the present pre-processing tool, respectively.
Concluding remarks are given in Section \ref{sec:conclusion}.
The source code of the present method 
is available in our open-source SPHinXsys library \cite{method--ZHANG/SPHinXsys, method--zhang2021} 
at \url{https://www.sphinxsys.org}.

%
%
\section{Preliminary work}
\label{sec:previous-work}
In this section, 
we briefly summarize the particle-relaxation method 
on generating body-fitted particle distributions 
for arbitrarily complex geometries 
and more details are referred to Ref. \cite{distribute--yujiezhu2021}. 
\subsection{level-set method and narrow-band technique}
\label{subsec:levelset-method} 
To represent the complex geometry, 
the level-set field  $\phi(x,y,z,t)$ is utilized by defining 
a signed distance function, so that the zero level-set contour 
\begin{equation}
\Gamma = \left\lbrace  \left(x, y, z \right)| \phi \left(x, y, z, t \right) = 0 \right\rbrace  .
\label{eq:level-set}
\end{equation}
represents the geometry surface, and the negative and positive 
level-set values are for the inside and outside regions respectively. 
The normal direction 
$\mathbf N = (n_x, n_y, n_z)^T$ can 
be evaluated by
\begin{equation}
	\mathbf N= \frac{\nabla \phi}{\left| \nabla \phi \right| }.
	\label{eq:normal}
\end{equation}
To discretize the level-set field,
a Cartesian background mesh is used in the  whole computational domain 
and the level-set value $\phi$ of each mesh cell is defined 
by the distance from the cell center to the geometry surface.
Subsequently, the level-set field can be constructed by 
parsing CAD data with proper in-house function or parser provided 
by open source library, for example Simbody 
\cite{method--sherman2011} and Boost libraries.  

A typical way to reduce the computational effort for level-set 
related operations is the narrow-band method 
\cite{narrow_band--adalsteinsson1995fast, narrow_band--gomez2005reinitialization, narrow_band--ye2012multigrid},
in which the computation effort is restricted to the near-interface band.
Here, the narrow-band technique similar to Refs.
\cite{narrow_band--gomez2005reinitialization, MR-mesh-data--han2014} is used 
with memory-pool data package technique \cite{MR-mesh-data--han2014} 
for further accelerating the level-set related operations.  

\begin{figure}
	\centering
	\begin{subfigure}[b]{0.49\textwidth}
		\centering
		\includegraphics[trim = 0cm 0cm 0cm 0cm, clip, width=0.9\textwidth]{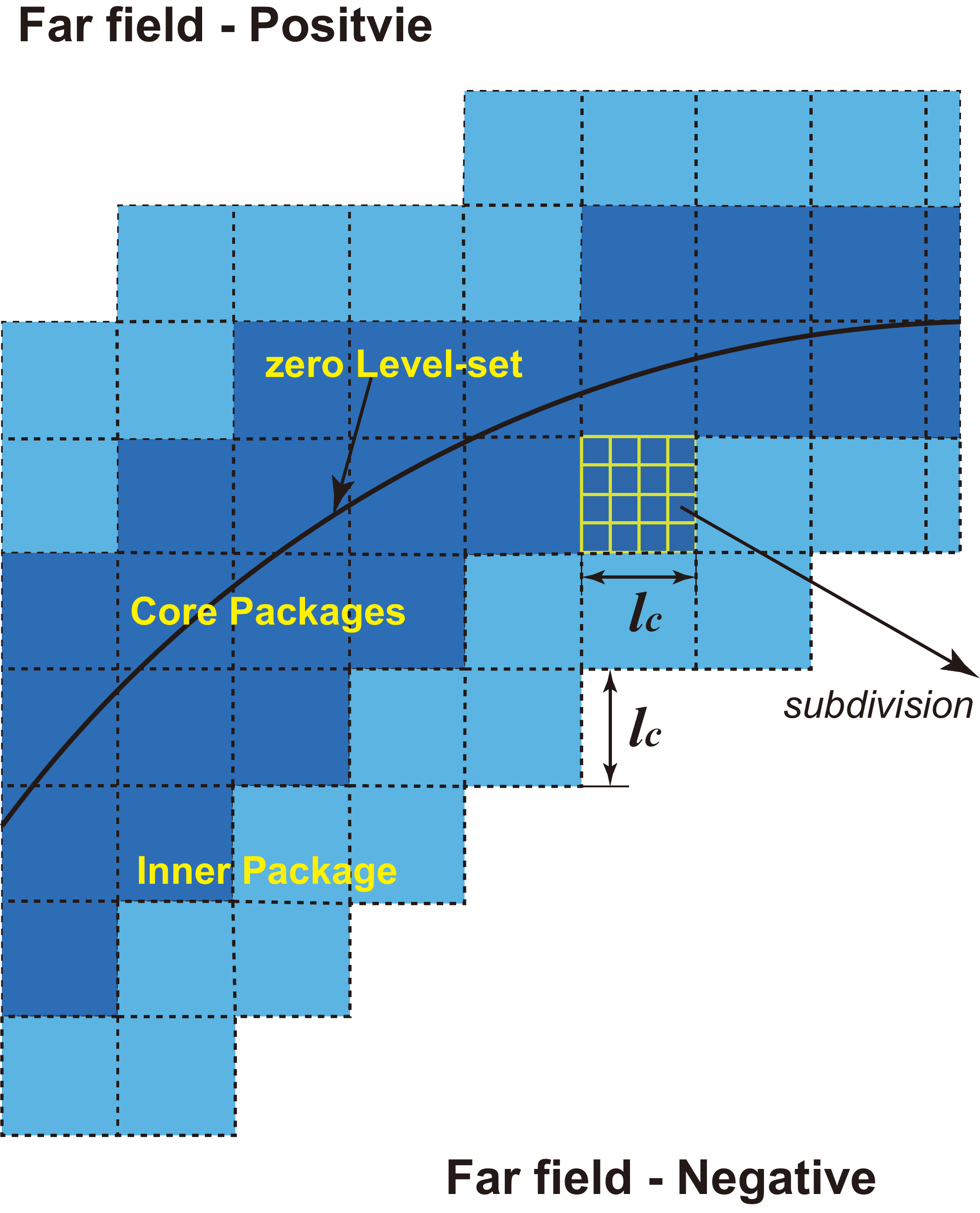}
		\caption{}
		\label{figs:narrow band (a)}
	\end{subfigure}
	\begin{subfigure}[b]{0.49\textwidth}
		\centering
		\includegraphics[trim = 0cm 0cm 0cm 0cm, clip, width=0.9\textwidth]{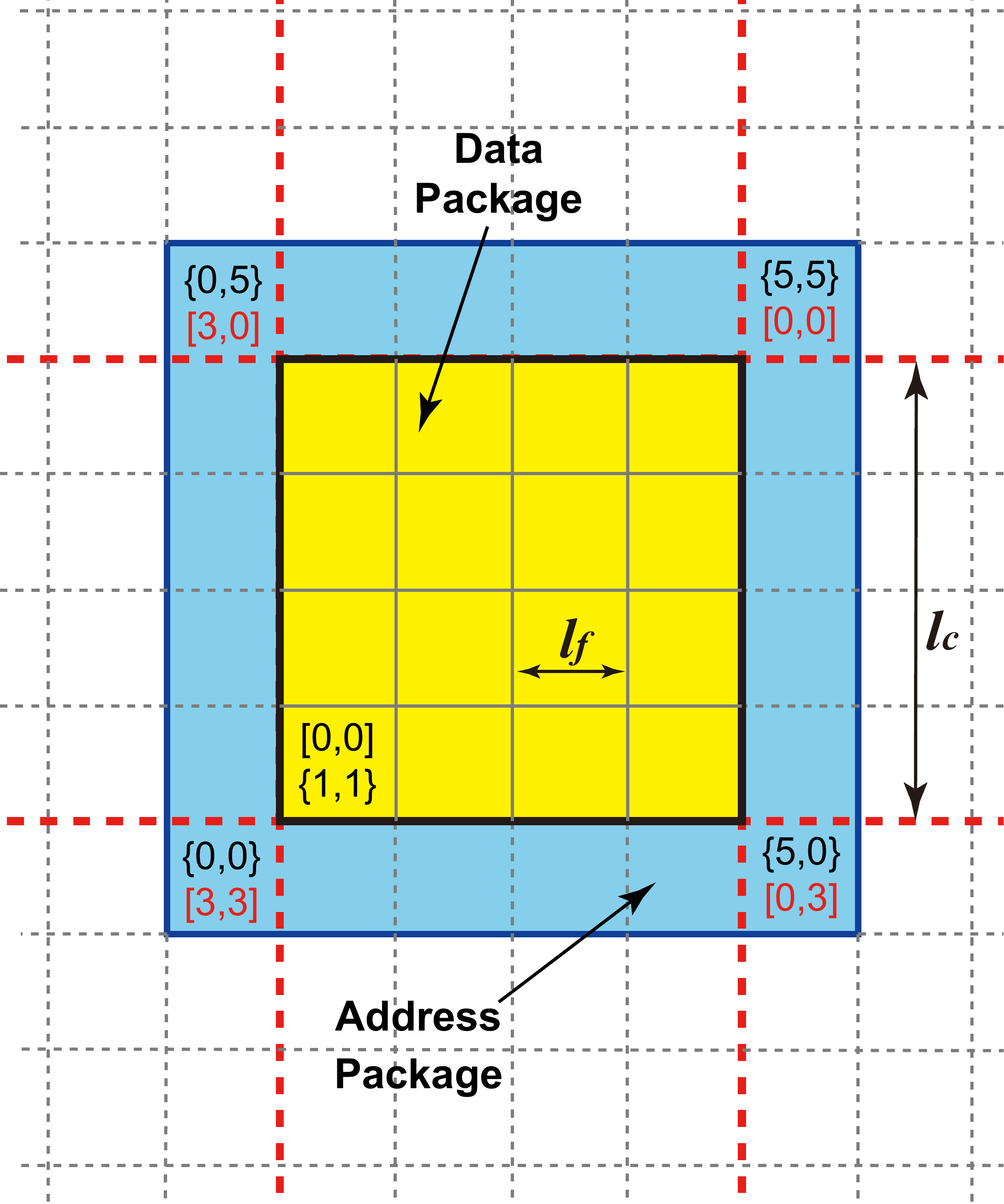}
		\caption{}
		\label{figs:package structure (b)}
	\end{subfigure}
	\caption{level-set field with narrow band data storage: 
	(a) `inner packages'(light blue cells) and `core packages'(dark blue cells)
	around geometry surface, 
		(b) structure of data-package (yellow cells) and address-package (blue cells). 
		The square brackets indicate a data sequence in the `data package'.
		The curly brackets indicate the address sequence in the `address package'.
		All the black indexes are belong to the same `data package' and its
		`address package', while the red indexes are from its neighbor `data package'. }
	\label{figs:data-structure}
\end{figure}
Fig. \ref{figs:data-structure} gives a detailed description of 
the narrow-band and data package settings. 
Here, the narrow-band region near 
the geometry surface is further divided into the core region and the inner region 
(the inner region contains the core region) 
represented by mesh cells with a coarse cell spacing $l_c$. Each mesh cell in the 
narrow-band region has been subdivided into $4 \times4$ (in 2D) data grids with the
fine cell spacing $l_f$ equals to a quarter of $l_c$ and named as data package. 
For every data package, 
there is an overlapped address package with one cell spacing $l_f$ over the 
data package in each direction for parallel operation and 
to improve the memory efficiency. 
For the narrow-band region, the level-set value is computed 
as the distance between the cells' center to the geometry surface.
While for the remaining area named as `far field' in our method, 
only two fixed values 
${-4 l_c }$ and ${  4 l_c }$ are assigned with their memory addresses to each cell  
depending on whether the `far field' cell is within the geometry or not.
That means, in the narrow-band method, there are two levels of mesh, 
one covers the whole computational area with the coarse cell spacing $l_c$, 
the other is only employed in the narrow-band region with fine cell spacing $l_f$. 
The detailed implementation of our narrow-band and package techniques please 
refer to the Algorithm \ref{alg:a}. 

\begin{algorithm}[htb!]
	Setup parameters and initialize the computation\;
	Read and parse the polygon mesh of a specific geometry from CAD files\;
	Divide calculation domain by a coarse cell spacing $l_c$ into each $C_{i,j}$\;
	Initialize the positive and negative far field cell $C_p = 4 l_c$ and $C_n = -4 l_c$\;
	\ForAll{$C_{i,j}$}{
    	Detect the distance $D_{i,j}$ of each coarse cell $C_{i,j}$ to $\Gamma_0$\;
    	\uIf{$D_{i,j} <= l_c$}{
    	Mark $C_{i,j}$ as ${Core\_package}$ and ${Inner\_package}$\;
    	Divide $C_{i,j}$ by a fine cell spacing $l_f$ into $4\times4$ ${data\_cells}$ $d_{m,n}$\;
    	Initialize ${Core\_package}$ with $\phi_{m,n}$ and its address\;}
    	\uElseIf{$C_{i,j}|\exists C_{i+i_0,j+j_0} \in \mbox{Core\_packages},\land (\forall i_0,j_0\in \{-1,0,1\}) $}{
    	Mark $C_{i,j}$ as ${Inner\_package}$\;
    	Divide $C_{i,j}$ by a fine cell spacing $l_f$ into $4\times4$ ${data\_cells}$ $d_{m,n}$\;
    	Initialize ${Inner\_package}$ with $\phi_{m,n}$ and its address\;}
    	\uElse {
    	Point the address of $C_{i,j}$ to $C_p$ or $C_n$ depends on the sign of its $D_{i,j}$\;}
    	Link the address of each $C_{i,j}$ of all ${Inner\_packages}$\;
    }
    \ForAll{$Inner\_packages$}{
    Initialize the normal direction of each $d_{m,n}$ by Eq.\ref{eq:normal}\;
    Initialize the $I_{i,j}$ of each $d_{m,n}$ by 
    Eq.\ref{eq:kernel gradient from levelset contribution}\;
    }
	Terminate the computation.
	\caption{Algorithm for the construction of the level-set field with narrow-band and
	package techniques (in 2D)}
	\label{alg:a}
\end{algorithm}
%

\subsection{Physics-driven particle relaxation} \label{subsec:physics-driven relaxation} 
Isotropic and body-fitted particle distribution can be obtained by 
implementing a physics-driven relaxation process 
with a level-set based surface particle bounding \cite{distribute--yujiezhu2021}. 
In the physics-driven relaxation process, 
the particle advection is governed by 
\begin{equation} \label{eq:momentum}
\frac{\text{d}\mathbf{v}}{\text{d}t} = \mathbf{F}_p,
\end{equation}
where $\mathbf{v}$ is the advection velocity, 
$\mathbf{F}_p$ denotes the accelerations due to the repulsive pressure force 
and $\frac{\text{d}\left(\bullet \right) }{\text{d} t}=\frac{\partial \left(\bullet \right)}{\partial t} + \mathbf{v} \cdot \nabla \left(\bullet \right)$ 
stands for the material derivative. 
Following Refs. \cite{distribute--yujiezhu2021, method--litvinov2015, Method--adami2013transport, Method--zhang2017generalized}, 
the pressure term in the right-hand-side of Eq.\ref{eq:momentum} can be calculated as 
\begin{equation}\label{eq:momentum-sph}
	\mathbf{F}_{p, a} = - \frac{2 p_{0}  V_a }{m_a}\sum_b \nabla_a W_{ab} V_b ,
\end{equation}
by applying a constant background pressure $p_0$ \cite{method--litvinov2015}. 
Here,
$m$ is the particle mass, 
$V$ the particle volume 
and $\nabla_a W_{ab}$ represents the gradient of the kernel function $W(|\mathbf r_{ab}|,h)$ 
with respect to particle $a$. 
Note that $\mathbf r_{ab} = \mathbf r_a - \mathbf r_b$ and $h$ represents the smoothing length. 
Following with the time-step size $\Delta t$ which is constrained by the body force criterion,
\begin{equation}\label{eq:time-step}
\Delta t \leq 0.25 \sqrt{\frac{h} {\left| \text d \mathbf{v} / \text d t\right|} },
\end{equation}
And the position updating method of particles as 
\begin{equation}\label{eq:update-position}
\mathbf{r^{n+1}}=\mathbf{r^n} + \text{d} \mathbf{r} =\mathbf{r^n} + \frac{1}{2} \mathbf{F}_p^n \Delta t^2.
\end{equation}
To take into account the lack of kernel support for near-surface particles, 
the particle position is modified by a surface bounding
\begin{equation}\label{eq:bounding}
	\mathbf r_a= 
	\begin{cases}
		\mathbf r_a -  \left( \phi_a + \frac{1}{2} \Delta x \right)  \mathbf N_a  \quad  &\phi_a \ge - \frac{1}{2} \Delta x \\
		\mathbf r_a  \quad  &\text{otherwise} 
	\end{cases},
\end{equation}
where $\Delta x$ denotes the particle spacing, 
and $\phi_a$ and $\mathbf N_a$ are the level-set value and the normal direction at 
the position of particle $a$, respectively. 
More details are referred to Ref. \cite{distribute--yujiezhu2021}.
%
%
\section{`Dirty' geometry cleaning}\label{sec:clean}
\begin{figure}
	\centering
	\centering
	\includegraphics[trim = 0cm 0cm 0cm 0cm, clip, width=0.9\textwidth]{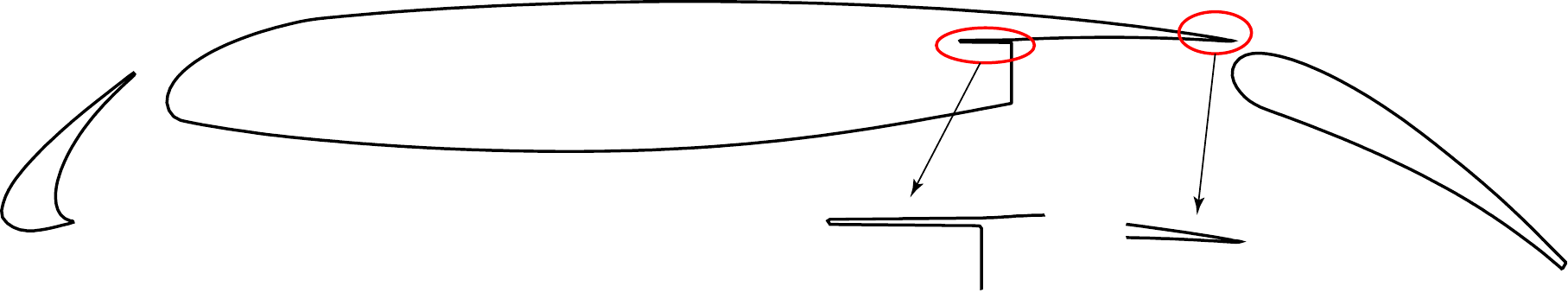}
	\caption{High-lift airfoil 30P30N with sharp trailing edge and a
		sharp slit at its second part.}
	\label{figs:airfoil-with-slit }
\end{figure}

\begin{figure}
	\centering
	\begin{subfigure}[b]{0.49\textwidth}
		\centering
		\includegraphics[trim = 0cm 0cm 0cm 0cm, clip, width=0.9\textwidth]{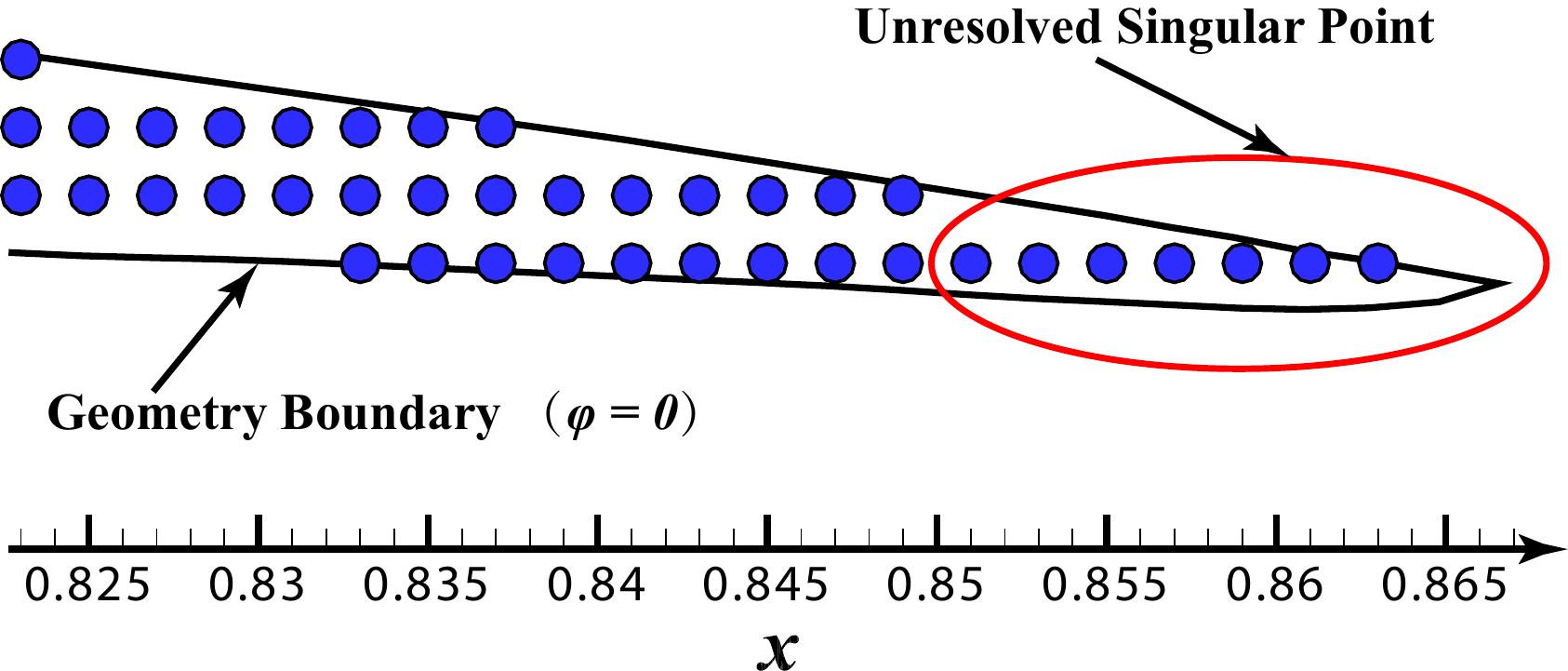}
		\caption{}
		\label{figs:Unstable relaxation in trailing edge (a)}
	\end{subfigure}
	\begin{subfigure}[b]{0.49\textwidth}
		\centering
		\includegraphics[trim = 0cm 0cm 0cm 0cm, clip, width=0.9\textwidth]{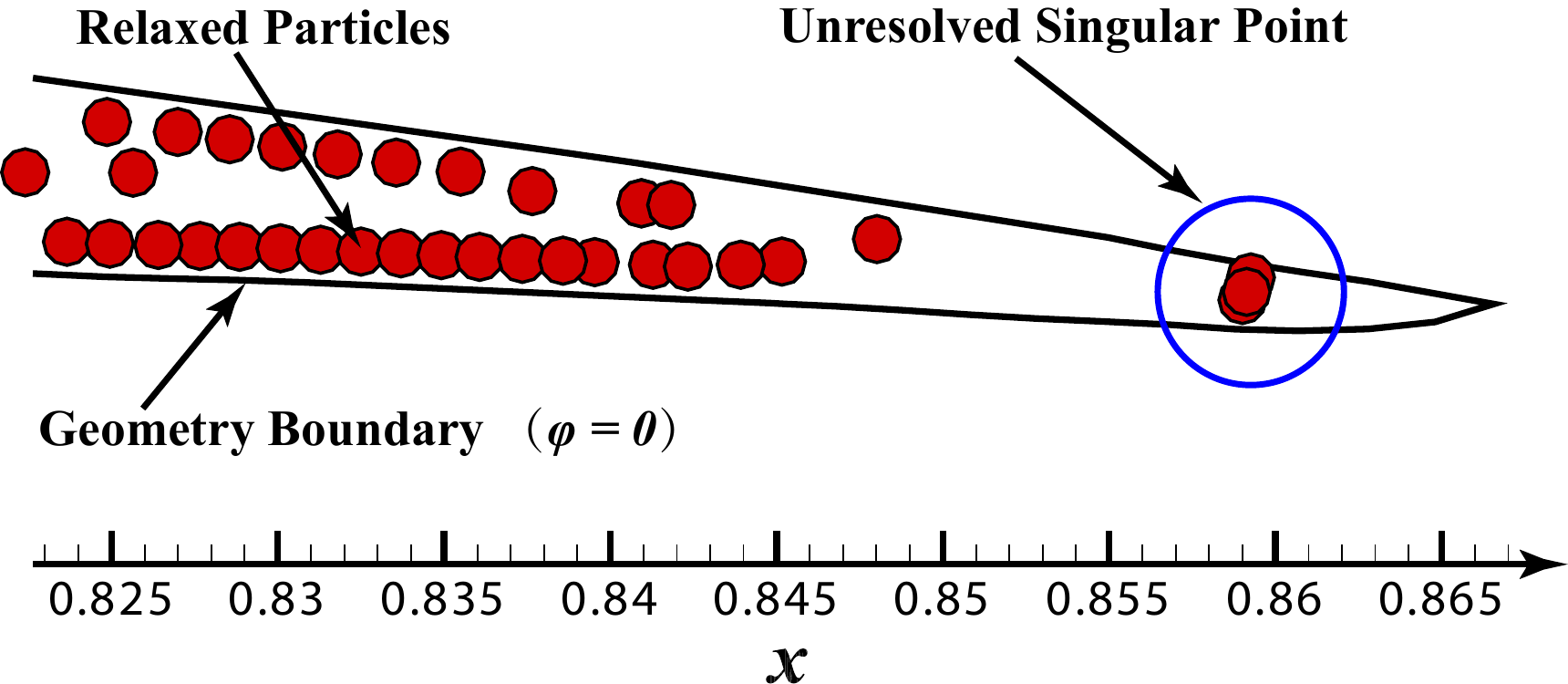}
		\caption{}
		\label{figs:Unstable relaxation in trailing edge (b)}
	\end{subfigure}
	\caption{Unstable particle generation process caused by unsolved small structure.
	(a) Mono-layer particles in trailing edge; (b) Unstable relaxation process.}
	\label{figs:Unstable particle generation process caused by unsolved thin structure}
\end{figure}

Here, we use a defective geometry of a high-lift airfoil 30P30N 
(shown in Fig. \ref{figs:airfoil-with-slit }) 
to illustrate the issue of `dirty' geometry.
The airfoil has a typical sharp trailing edge and a small slit 
which is considered as a defect in its second part. 
These parts are left unresolved even very high resolution is applied.
As shown in Fig. \ref{figs:Unstable relaxation in trailing edge (a)},
although the grid resolution used to generate lattice particle distribution is high 
enough compared to the chord length of the airfoil (0.002 to 1), 
several mono-layer particles are generated at the end of the sharp trailing edge with lattice distribution. 
  
These mono-layer particles leads to unresolved singularity 
and can induce numerical instability during the relaxation process as 
shown in Fig. \ref{figs:Unstable relaxation in trailing edge (b)},
and failures of numerical simulations based on such particle distribution. 
Although increasing the resolution to fit the sharp non-resolved geometry 
could address this issue to some extent, 
excessive computational efforts are also inevitable.
A possible solution is to clean up these non-resolved parts,
as their impact on the simulation results 
under a given resolution very often is negligible.

%

\subsection{Identify non-resolved small structures}
\label{subsec:clean-mark}
\begin{figure}
	\centering
	\centering
	\includegraphics[trim = 0cm 0cm 0cm 0cm, clip, width=0.9\textwidth]{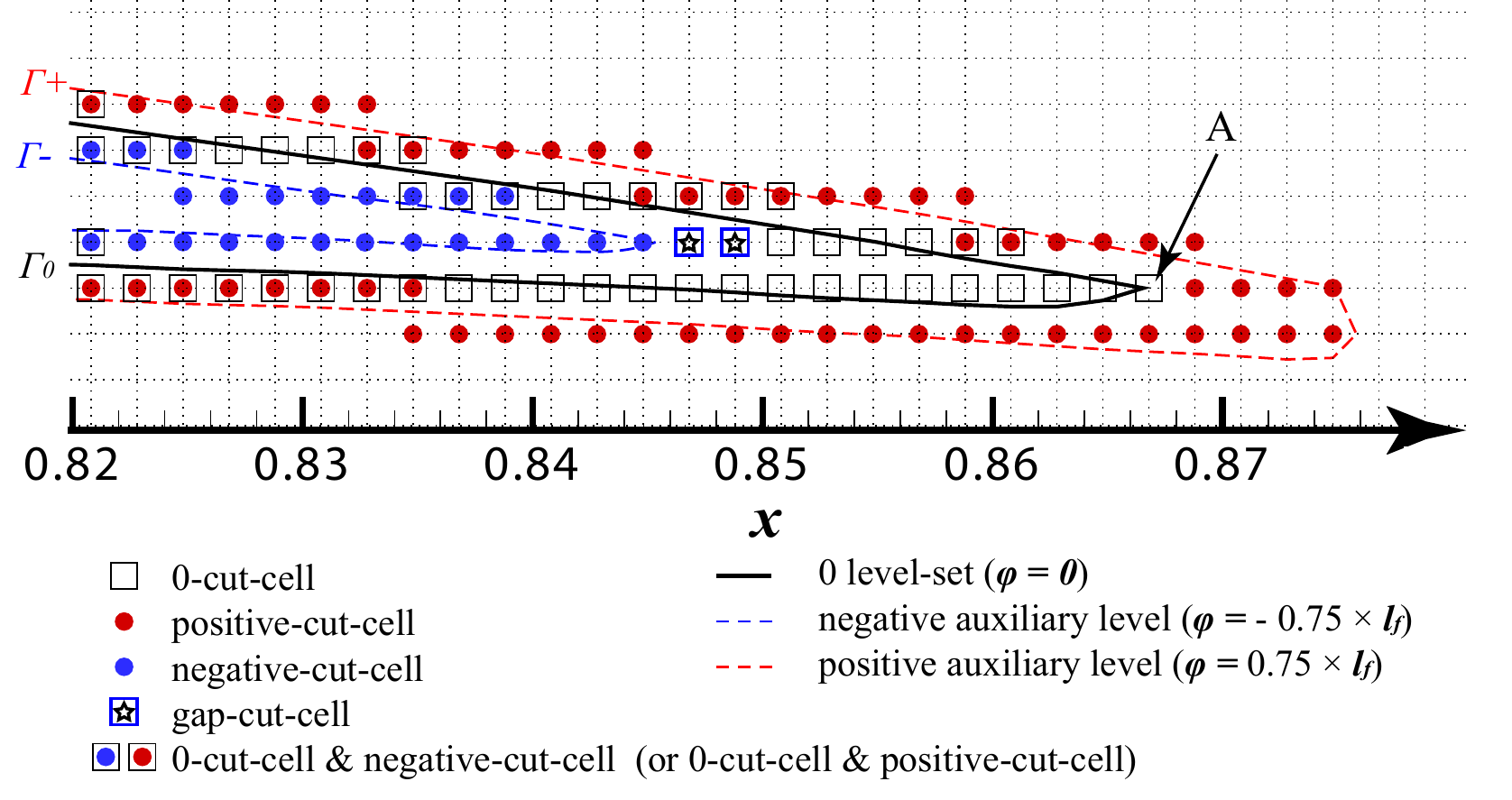}
	\caption{Mark non-resolved small structure. 
	The 0-cut-cell `A' has a 11 cell-spacing distance to nearest negative 
	auxiliary level $\Gamma_-$, which is already far beyond its searching 
	region, a $5 \times 5$ mesh cells area centered on `A'. Note that the 
	background mesh are connected by the center of each cell.}
	\label{figs:levelset mark}
\end{figure}

Inspired by the scale separation method in Luo's 
work\cite{Method--luo2016}, an approach to finding the non-resolved structures 
of a geometry surface by detecting whether the mesh cell across zero 
level-set or an auxiliary level is employed in this paper 
(shown in Fig. \ref{figs:levelset mark}).   
In order to distinguish different mesh cells, 
we call the mesh cells cut through by geometry surface $\Gamma_0$ the `0-cut-cells', 
represented by `$C_0$'. 
\begin{equation} \label{eq:0-cut-cell}
 C_0 = \{C_{i,j}|\exists \phi_{i\pm1/2,j\pm1/2} \cdot \phi_{i\pm1/2,j\pm1/2} < 0 \},
\end{equation}
where $C_{i,j}$ is a mesh cell indexed by $[i,j]$ and $\phi_{i\pm1/2, j\pm1/2}$ 
is the level-set value at the corner of $C_{i,j}$, 
can be obtained by the following interpolation:
\begin{equation} \label{eq:corner phi}
\begin{aligned}
\phi_{i+1/2,j+1/2} = \frac{1}{4}(\phi_{i,j}+\phi_{i+1,j}+\phi_{i,j+1}+\phi_{i+1,j+1})\\
\phi_{i+1/2,j-1/2} = \frac{1}{4}(\phi_{i,j}+\phi_{i+1,j}+\phi_{i,j-1}+\phi_{i+1,j-1})\\
\phi_{i-1/2,j+1/2} = \frac{1}{4}(\phi_{i,j}+\phi_{i-1,j}+\phi_{i,j+1}+\phi_{i-1,j+1})\\
\phi_{i-1/2,j-1/2} = \frac{1}{4}(\phi_{i,j}+\phi_{i-1,j}+\phi_{i,j-1}+\phi_{i-1,j-1})\\
\end{aligned}.
\end{equation}
Then the mesh cells cut through by positive auxiliary level 
$\Gamma_+$ and negative auxiliary level $\Gamma_-$ 
are named as the `positive-cut-cells' and the `negative-cut-cells', 
respectively. Which yield
\begin{equation} \label{eq:positive-cut-cell}
 C_{+\epsilon} = \{C_{i,j}|\exists (\phi_{i\pm1/2,j\pm1/2} - \epsilon) \cdot (\phi_{i\pm1/2,j\pm1/2} - \epsilon) < 0 \},
\end{equation}
\begin{equation} \label{eq:negative-cut-cell}
 C_{-\epsilon} = \{C_{i,j}|\exists (\phi_{i\pm1/2,j\pm1/2} + \epsilon) \cdot (\phi_{i\pm1/2,j\pm1/2} + \epsilon) < 0 \},
\end{equation}
where $\epsilon$ is $0.75l_f$ as in Ref. \cite{method--han2015}.

Different from Luo's work \cite{Method--luo2016}, 
mesh cells which have the value of level-set between $0$ and $-\epsilon$ ( $0>\phi>-\epsilon$) 
as well as those level-set value between $0$ and $+\epsilon$ ( $+\epsilon>\phi>0$) 
must be found out and marked as `gap-cut-cell'. This step is necessary, 
and the detailed explanation is given in section \ref{subsec:clean-re-distance}. 
According to the topological consistency, the mesh cells with a non-resolved 
geometry surface are those from `$C_0$' and `$C_g$' but have only positive or 
negative auxiliary cells as their neighbor cells. Note that the cut cell identifying 
operations only need to be executed in `core packages'.

\begin{equation} \label{eq:non-resolved cut cell}
\begin{aligned}
 C_{non-re} =&\{C_{i,j}|C_{i,j} \in (C_0 \cup C_g) \land (\forall i_0,j_0\in \{-1,0,1\},\\ &C_{i+i_0,j+j_0} \notin (C_{+\epsilon} \cup  C_{-\epsilon}) )\}
\end{aligned}
\end{equation}
For better understand, all the cut-cells marked by `$C_0$', `$C_{+\epsilon}$', 
`$C_{-\epsilon}$' and `$C_g$' have a unified title `Interface ID' in Algorithm \ref{alg:b}.

\subsection{level-set re-distance and reinitialize }
\label{subsec:clean-re-distance}
To reconstruct the non-resolved geometry surface, 
one should modify the level-set value of those identified mesh cells with non-resolved 
segments. Inspired by the interface reconstruction method in Ref. \cite{Method--luo2016}, 
to facilitate the implementation, the non-resolved cut-cells $C_{non-re}$ should 
be separated into two types, $C_{non+}$, which has no neighbors belonging 
to $C_{+\epsilon}$ within one cell-spacing distance from it:
\begin{equation} \label{eq:non-resolved cut cell without positive neighbour}
 C_{non+} =\{C_{i,j}|C_{i,j} \in (C_0 \cup C_g) \land (\forall i_0,j_0\in \{-1,0,1\}, C_{i+i_0,j+j_0} \notin C_{+\epsilon})\},
\end{equation}
and $C_{non-}$, which has no neighbours belonging to $C_{-\epsilon}$ within one cell-spacing distance from it:
\begin{equation} \label{eq:non-resolved cut cell without negative neighbour}
 C_{non-} =\{C_{i,j}|C_{i,j} \in (C_0 \cup C_g) \land (\forall i_0,j_0\in \{-1,0,1\}, C_{i+i_0,j+j_0} \notin C_{-\epsilon})\}.
\end{equation}

According to Ref. \cite{Method--luo2016}, 
the level-set value of each re-distanced cut-cell in $C_{non+}$ and $C_{non-}$ is 
replaced by an estimated value of their distance to $\Gamma_+$ and $\Gamma_-$, respectively.
However, as the geometry is relatively too sharp under a given resolution 
in many cases (Fig. \ref{figs:levelset mark}), 
the $\Gamma_+$ or $\Gamma_-$ is far from geometry boundary. 
Considering that a large normal distance from the re-distanced cut-cell 
to a far-away auxiliary level is meaningless for resetting the level-set value, 
thus a searching region of $5 \times 5$ mesh cells area (in 2D) centered on 
the considered cut-cell is set by taking into account the computational efficiency. 
Meanwhile, a maximum replacement distance $D_{limit} = 3 \times l_f$ should 
also be given to prevent there being no corresponding auxiliary level in the search range, 
where the $3 \times l_f$ stands for a half cell-spacing beyond the searching radius, 
which means there is no auxiliary level in the searching region.

Then considering a cut cell `$A$' in $C_{non+}$, 
one can get a normal ray pointing from the center of `$A$' to the auxiliary 
level $\Gamma_+$ with an intersection point. The distance $D$ between the cell 
center to the intersection point can be calculated by
\begin{equation} \label{eq:distance between non-resolved cut cell to auxiliary band}
 D =\sqrt{(D_{cell-i} + \phi N_x)^2 + (D_{cell-j} + \phi N_y)^2},
\end{equation}
where the $(D_{cell-i},D_{cell-j})$ are the cell distance in $x$ and $y$ direction 
between the considered cut cell `$A$' in $C_{non+}$ and a cell `$P$' from auxiliary 
level $C_{+\epsilon}$ in the searching region respectively. 
$(N_x, N_y)$ denotes the unit normal vector and $\phi$ is the level-set value of cell `$P$'. 
Then, the replaced level-set value for cut cell `$A$' is:
\begin{equation} \label{eq:replaced phi value for non-resolved cut cell}
 \phi_{replace} = -\min(D_{min}, D_{limit}),
\end{equation}
where $D_{min}$ is the minimum value from 
Eq.\ref{eq:distance between non-resolved cut cell to auxiliary band}. 
When there is no auxiliary level in the searching region,
$D$ is considered infinite. Meanwhile, 
the replacement of the level-set value for a cut cell `$B$' in $C_{non-}$ can 
be obtained in the same way by only substituting $-\phi$ for $\phi$ in 
Eq.\ref{eq:distance between non-resolved cut cell to auxiliary band} and the 
negative sign times the right side of Eq.\ref{eq:replaced phi value for non-resolved cut cell}. 
Note that all the re-distance operations only need to be executed in `core packages'.

Here is an additional explanation of why the `gap-cut-cell' is marked in 
the section \ref{subsec:clean-mark}. Considering some fairly sharp and narrow geometries,
like the sharp trailing edge in Fig.\ref{figs:levelset mark},
the interval in $x$ direction between $\Gamma_0$ and $\Gamma_+$ or $\Gamma_-$ 
is relatively large compared with the distance in the vertical direction 
between two layers of $\Gamma_+$ or $\Gamma_-$ in the sharp corner.
Thus, the space between the new geometry surface and the auxiliary level ($\Gamma_-$ or $\Gamma_+$)
may be greater than one cell-spacing even replacing the level-set value 
of the non-resolved cut cell in the corner by $D_{limit}$.
As shown in Fig. \ref{figs:levelset mark},
the level-set value of the two cells in `$C_g$' will not be changed during the re-distance 
process if they are not marked as non-resolved cut cells.
Even though their right neighbor cells are the endpoints of the geometry surface, 
there is still a two-cell-spacing distance between $\Gamma_0$ and $\Gamma_-$, 
which does not conform to the topological consistency.

Since the level-set values may not be smooth after the re-distance operation, 
a re-initialization process is performed within the scope of `inner packages' 
to achieve a continuous distribution of $\phi$\ 
by the following equation \cite{method--sussman1998}
\begin{equation}\label{eq:re-initialized}
\phi_{\tau} + sgn(\phi)(\left| \nabla \phi \right|-1) = 0, \\
\end{equation}
where $\tau$ is a pseudo time and $sgn(\phi)$ represents a sign function to maintain 
the signed distance property of level-set function.

\begin{figure}
\centering
    \begin{subfigure}[b]{0.49\textwidth}
         \centering
         \includegraphics[trim = 0cm 0cm 0cm 0cm, clip,width=\textwidth]{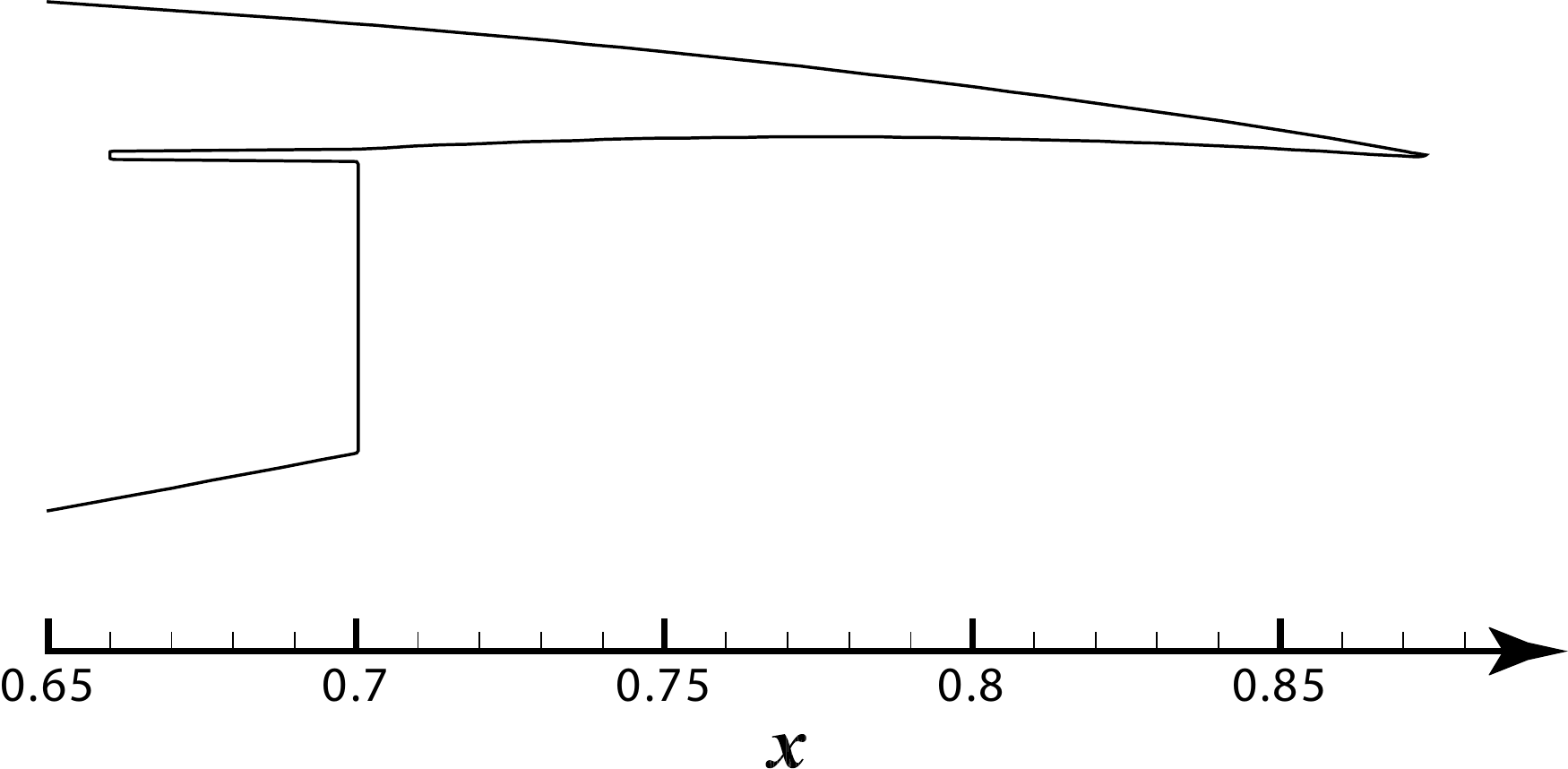}
         \caption{}
		\label{figs:re-constructed(a)}
     \end{subfigure}
    \begin{subfigure}[b]{0.49\textwidth}
         \centering
         \includegraphics[trim = 0cm 0cm 0cm 0cm, clip,width=\textwidth]{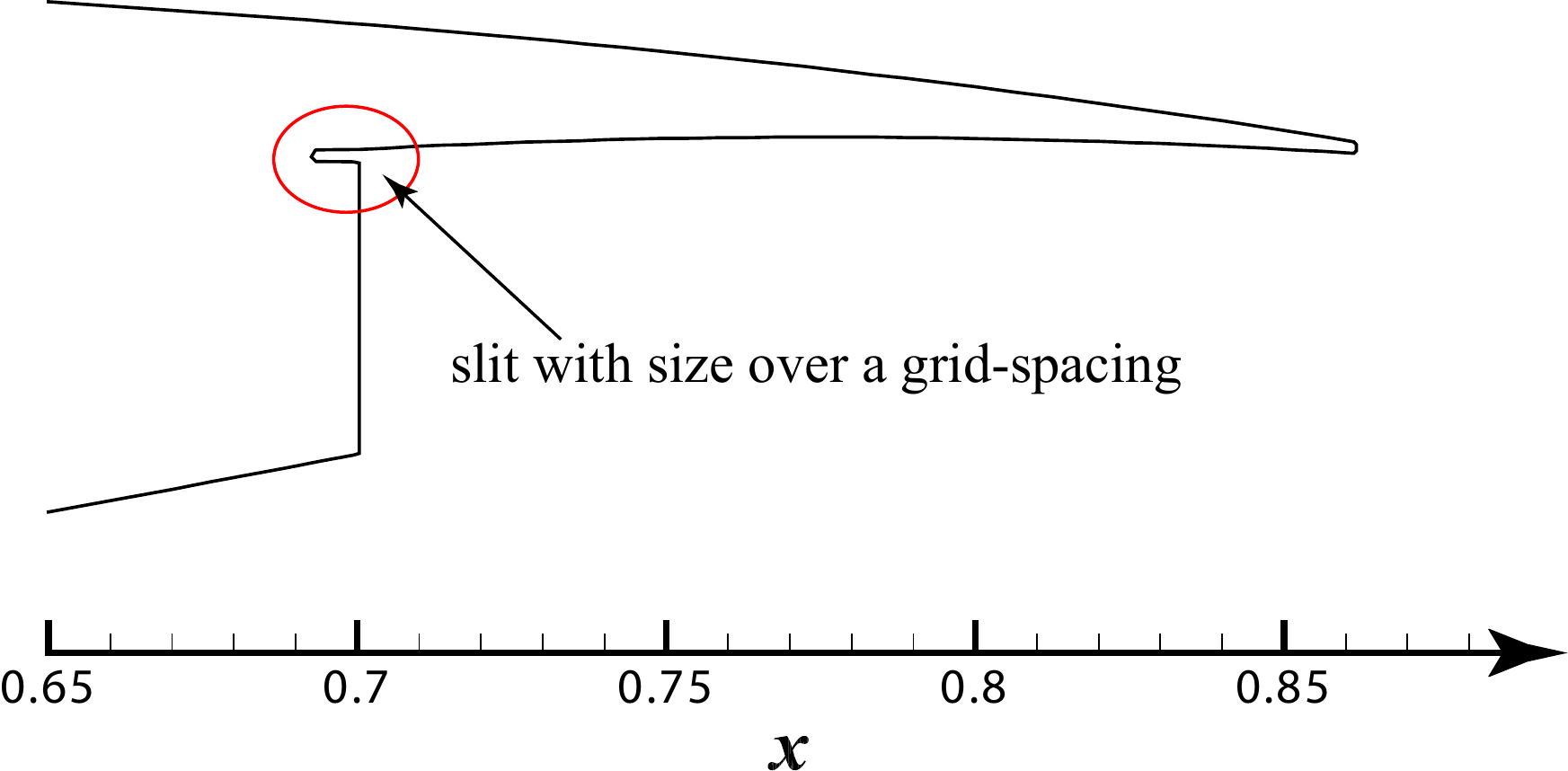}
         \caption{}
		\label{figs:re-constructed(b)}
     \end{subfigure}
     
    \begin{subfigure}[b]{0.49\textwidth}
        \centering
         \includegraphics[trim = 0cm 0cm 0cm 0cm, clip,width=\textwidth]{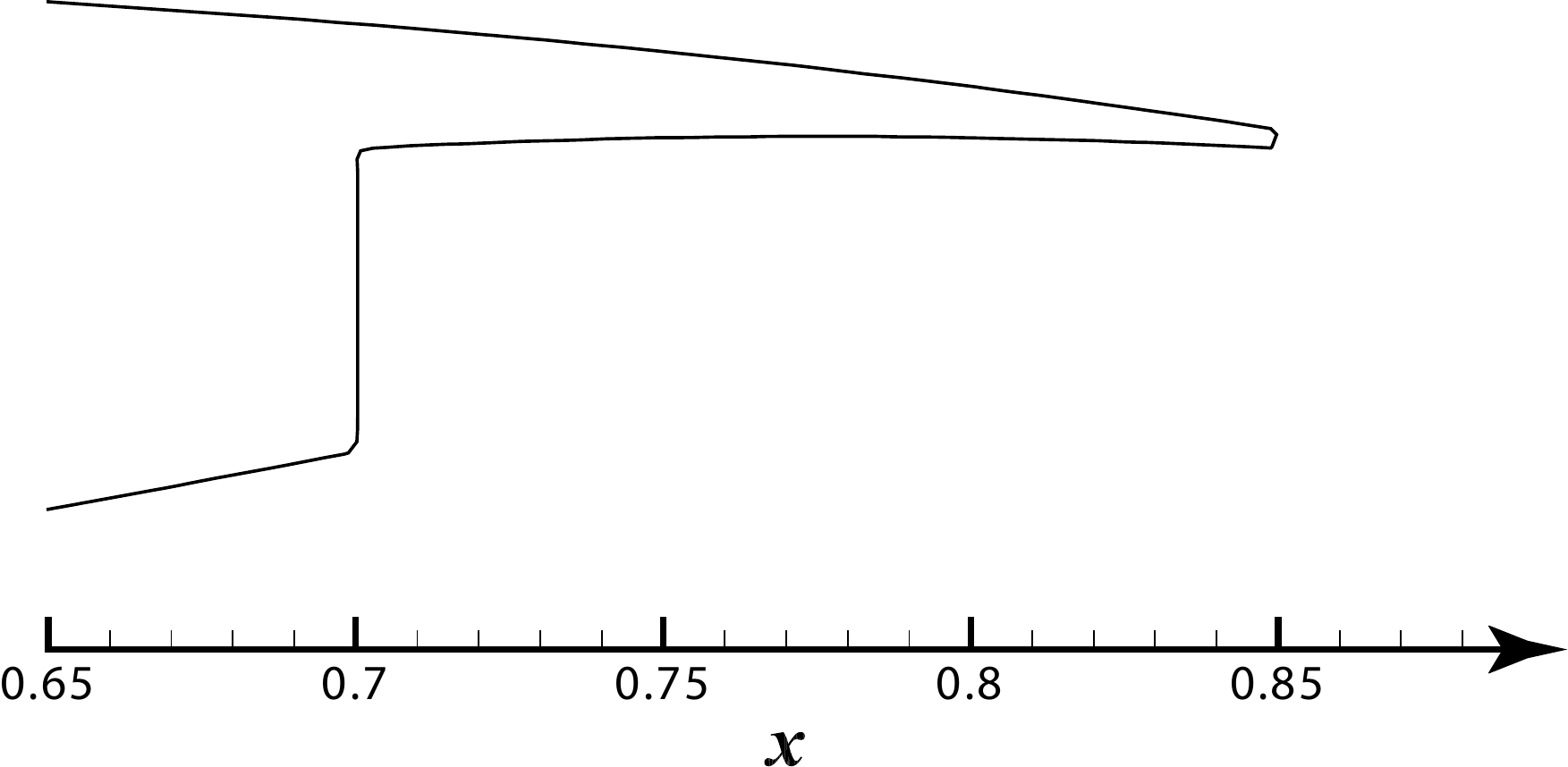}
         \caption{}
		\label{figs:re-constructed(c)}
    \end{subfigure}
    \begin{subfigure}[b]{0.49\textwidth}
        \centering
         \includegraphics[trim = 0cm 0cm 0cm 0cm, clip,width=\textwidth]{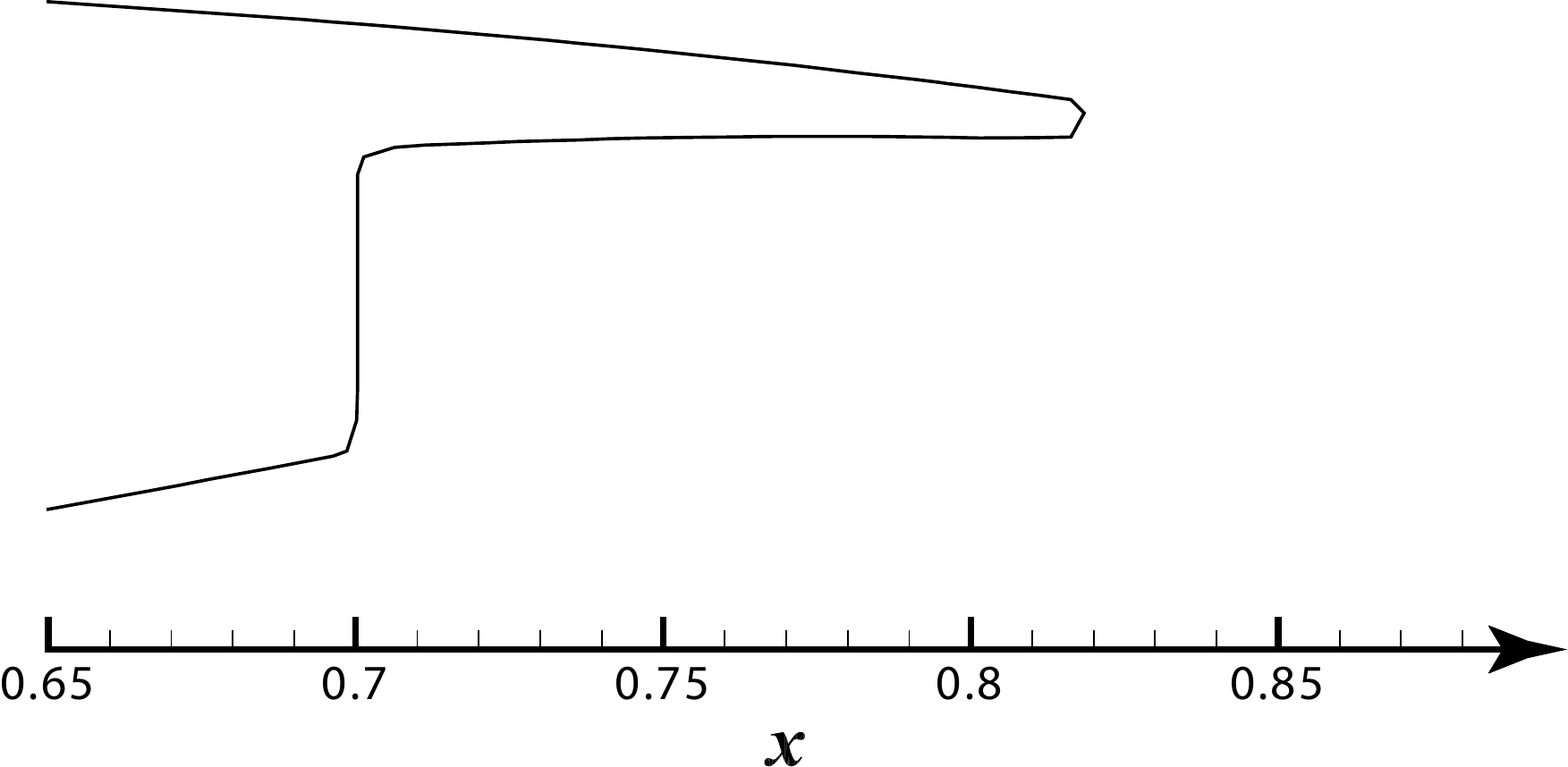}
         \caption{}
		\label{figs:re-constructed(d)}
    \end{subfigure}
 \caption{Airfoil surface reconstructed under different resolution. A small slit 
 is added on the second part of the airfoil mimics a defect in the geometric model. 
 (a) Original geometry surface; (b) Reconstructed with resolution 0.001; 
 (c) Reconstructed with resolution 0.002; (d) Reconstructed with resolution 0.005.}
 \label{figs:airfoil-boundary-re-constructed}
\end{figure}

Fig. \ref{figs:airfoil-boundary-re-constructed} shows the reconstructed geometry 
surface under different resolutions. A small slit is added to the second part of the 
high-lift airfoil 30P30N to mimic a defect in the geometric model. 
Together with the original sharp trailing 
edge (shown in Fig. \ref{figs:re-constructed(a)}), these non-resolved geometry fragments 
under a given resolution need to be reconstructed before particle or mesh generation. 
At a relatively high resolution, the rear part of the small slit is not recognized as 
the non-resolved segment since the width of this part is beyond a cell-spacing, 
and there is a tiny slit left at the corner between the airfoil tail and the main body 
as shown in Fig. \ref{figs:re-constructed(b)}. However, the non-resolved sharp trailing 
edge is reconstructed with a smoothed filleted corner which has a minimum size larger 
or equal to two cell-spacing. With the decreasing of the resolution 
(from Fig. \ref{figs:re-constructed(b)} to Fig. \ref{figs:re-constructed(d)}),
more portion of the trailing edge is identified as a non-resolved segment and reconstructed.
When the small slit is completely cleaned up at the resolution of 0.002, 
no additional variation at this area of the geometric surface presents.
This can prove that the present reconstruction process will automatically 
stop as the topological consistency is satisfied.  
\begin{figure}
	\centering
	\includegraphics[trim = 0cm 0cm 0cm 0cm, clip, width=0.8\textwidth]{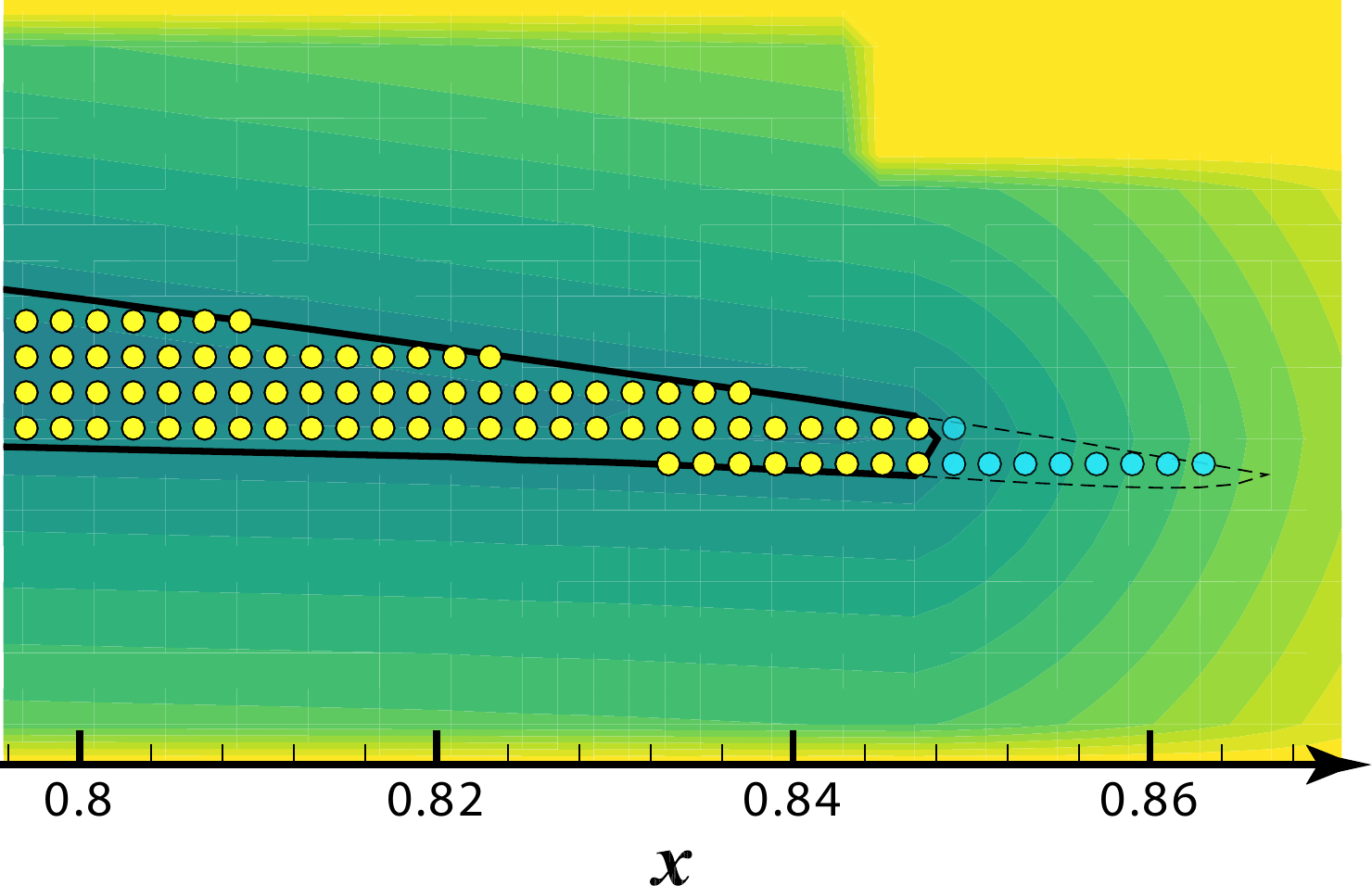}
	\caption{Lattice particles generated after clean the non-resolved structure. 
	The grey dots represent those unresolved singular points but vanished after the clean process.}
	\label{figs:lattice particles after clean }
\end{figure}

Figure \ref{figs:lattice particles after clean } shows the lattice particle distribution 
at the trailing edge of the airfoil after reconstructing the non-resolved geometric surface. 
It can be observed that the single layer unresolved singular points 
(shown as grey dots in Fig. \ref{figs:lattice particles after clean }) are vanished 
compared with Figure \ref{figs:Unstable relaxation in trailing edge (a)}. 
The two points on different layers at the far left of those removed particles are also included, 
which is due to the cells in `$C_g$' being recognized as unresolved fragments. 
The `dirty' geometry clean-up process is illustrated in Algorithm \ref{alg:b}. 

\begin{algorithm}[htb!]
	\uIf{Geometry needs to be clean}{
	\For{all $Core\_packages$}{
    	Mark $InterfaceID_{(m,n)}$ of every $d_{m,n}$\;
    	Find out the cut-cells $\in C_{non-re}$ \;
    	Re-distance the $\phi_{m,n}$ for cut-cells $\in C_{non+}$ and $\in C_{non-}$ 
    	respectively according to Eq.\ref{eq:distance between non-resolved cut cell to auxiliary band} 
    	and Eq.\ref{eq:replaced phi value for non-resolved cut cell}\;
    }
    \For{all $Inner\_packages$}{
    Reinitialize the $\phi_{m,n}$ of each $d_{m,n}$ by Eq.\ref{eq:re-initialized}\;
    Update the normal direction of each $d_{m,n}$ by Eq.\ref{eq:normal}\;
    Update the $I_{i,j}$ of each $d_{m,n}$ by 
    Eq.\ref{eq:kernel gradient from levelset contribution}\;
    }
	}
	Terminate the simulation.
	\caption{Algorithm for `dirty' geometry clean-up}
	\label{alg:b}
\end{algorithm}
%

%
%
\section{`Static confinement' boundary condition}\label{sec:static confinement}
As mentioned in the introduction,
the simple bounding method  \cite{distribute--yujiezhu2021} 
constraints the particles within the geometry using level-set value and the normal direction. 
But the particles 
near the geometry surface still 
have an incomplete support domain during the physics-driven relaxation process. 
This will not lead to a serious problem to those geometric surfaces with 
smooth shapes and small curvature.
\begin{figure}
	\centering
	\begin{subfigure}[b]{0.49\textwidth}
		\centering
		\includegraphics[trim = 0cm 0cm 0cm 0cm, clip, width=0.9\textwidth]{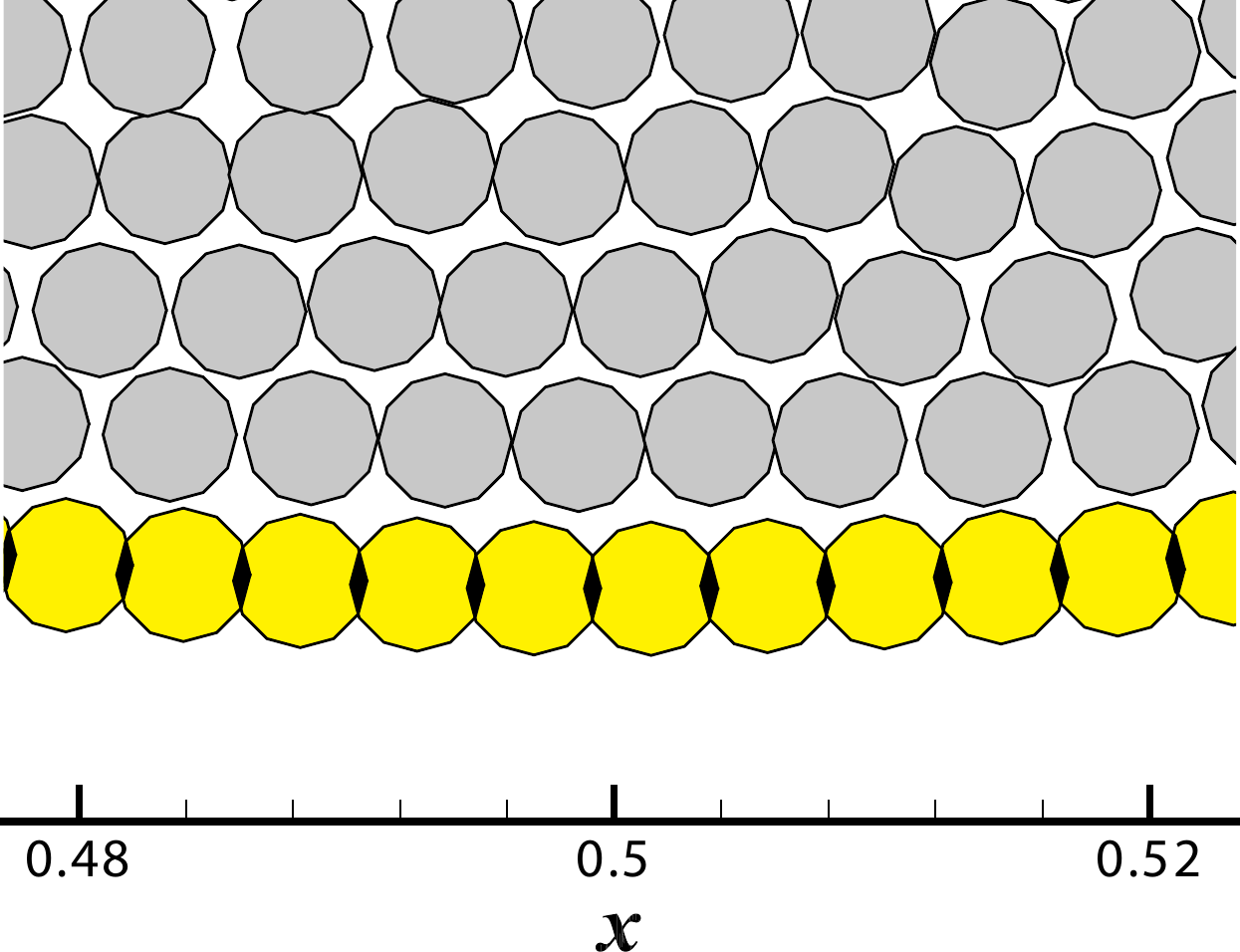}
		\caption{}
		\label{}
	\end{subfigure}
	\begin{subfigure}[b]{0.49\textwidth}
		\centering
		\includegraphics[trim = 0cm 0cm 0cm 0cm, clip, width=0.9\textwidth]{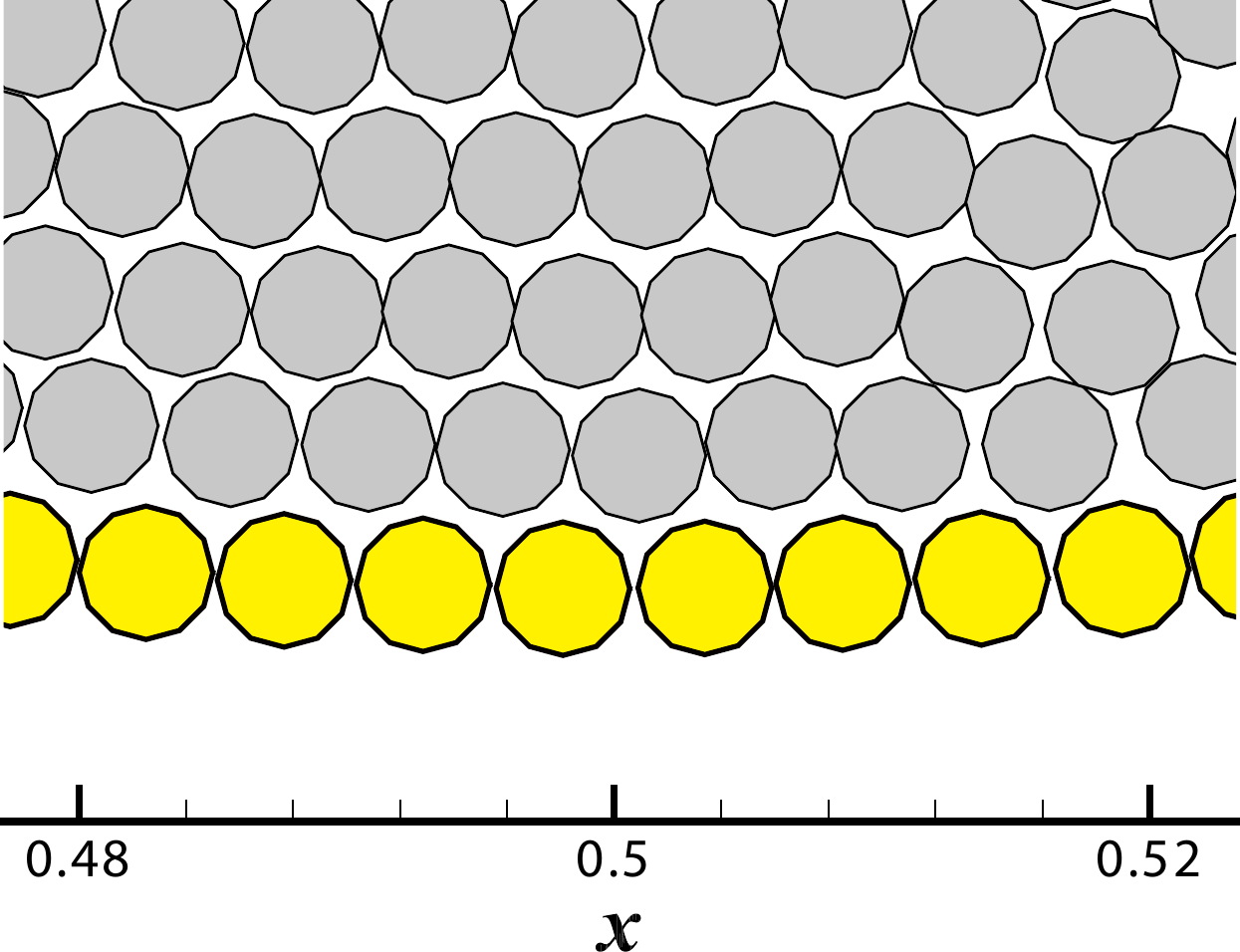}
		\caption{}
		\label{}
	\end{subfigure}
	\caption{Boundary particle distribution of a circle without `static confinement' 
		(a) and with `static confinement' (b). 
		Note that without `static confinement' boundary condition, 
		the (yellow) particle at the outer layer exhibit smaller distances than inner (gray) particles.}
	\label{figs:particle_distribution_circle}
\end{figure}
As shown in Fig.\ref{figs:particle_distribution_circle}, 
without the complete kernel support, 
although the relaxation process converges,
the boundary particles have a smaller particle spacing comparing with inner particles.
This will cause a slight inconsistency between particle
distribution and particle volume.
However, for those geometries with greater curvature and sharp features, 
the particle relaxation may not converge with incomplete kernel support.
\begin{figure}
	\centering
	\begin{subfigure}[b]{0.9\textwidth}
		\centering
		\includegraphics[trim = 0cm 0cm 0cm 0cm, clip, width=0.9\textwidth]{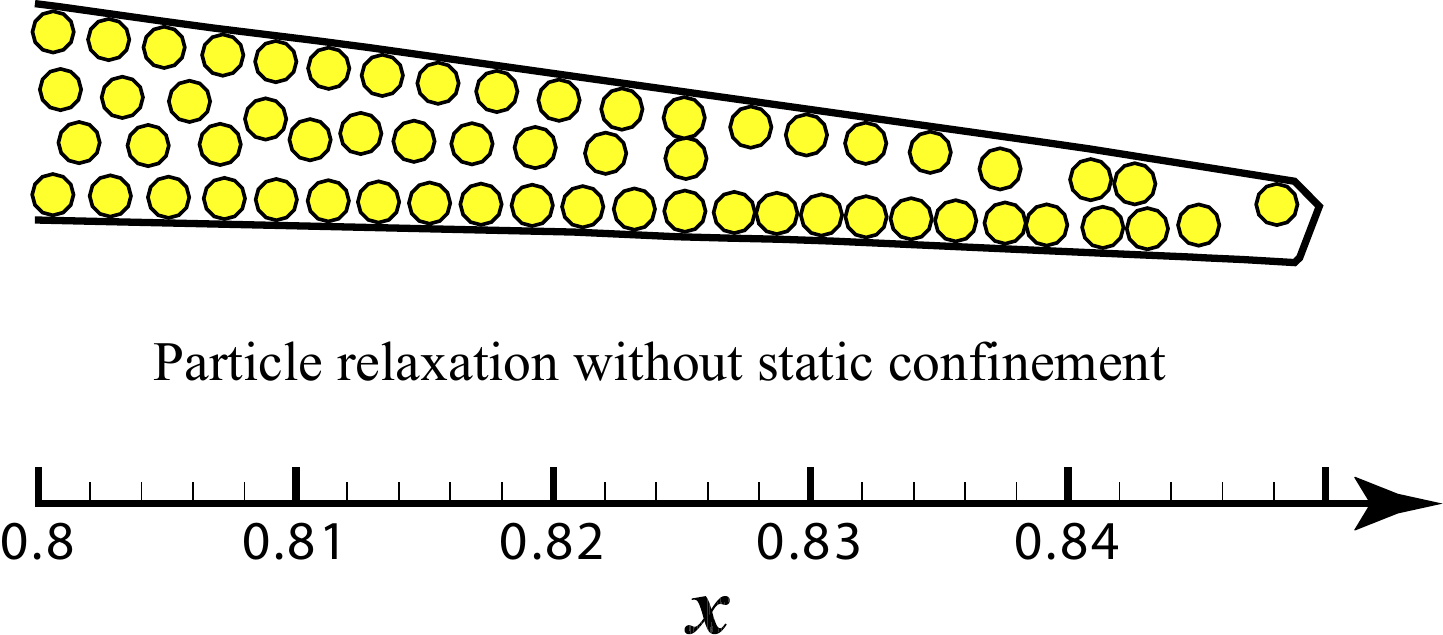}
		\caption{}
		\label{}
	\end{subfigure}
	
	\begin{subfigure}[b]{0.9\textwidth}
		\centering
		\includegraphics[trim = 0cm 0cm 0cm 0cm, clip, width=0.9\textwidth]{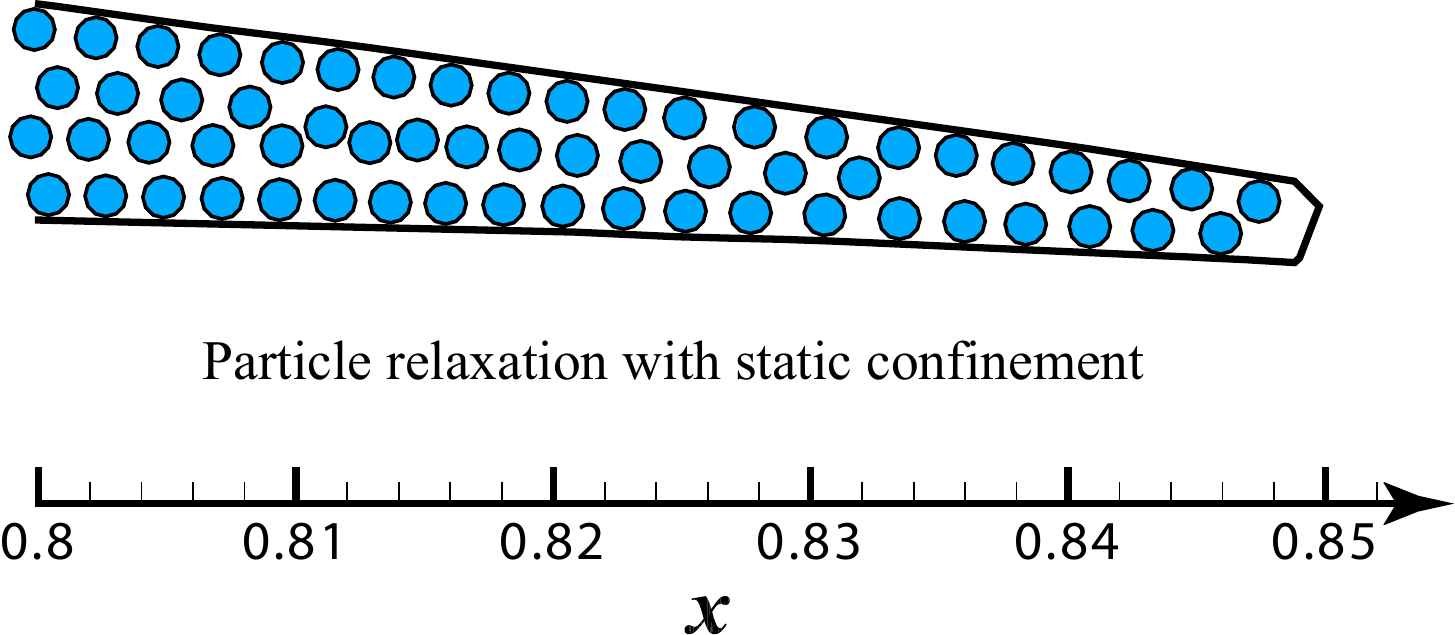}
		\caption{}
		\label{}
	\end{subfigure}
	\caption{Particle distribution at trailing edge without static confinement (a) and with static confinement (b) .}
	\label{figs:particle_distribution_airfoil}
\end{figure}
It is clearly shown in a snapshot of particle distribution in Fig. \ref{figs:particle_distribution_airfoil}a, 
where a few particles at the very end of the airfoil trailing edge are on the position 
where the particle relaxation does not converge, 
but with persistent cycling motion.  
In order to address the above-mentioned issue, 
here, we propose a method based on the level-set field to achieve full kernel support, 
which is denoted as `static confinement'. 

\begin{figure}
	\centering
	\includegraphics[trim = 0cm 0cm 0cm 0cm, clip, width=0.8\textwidth]{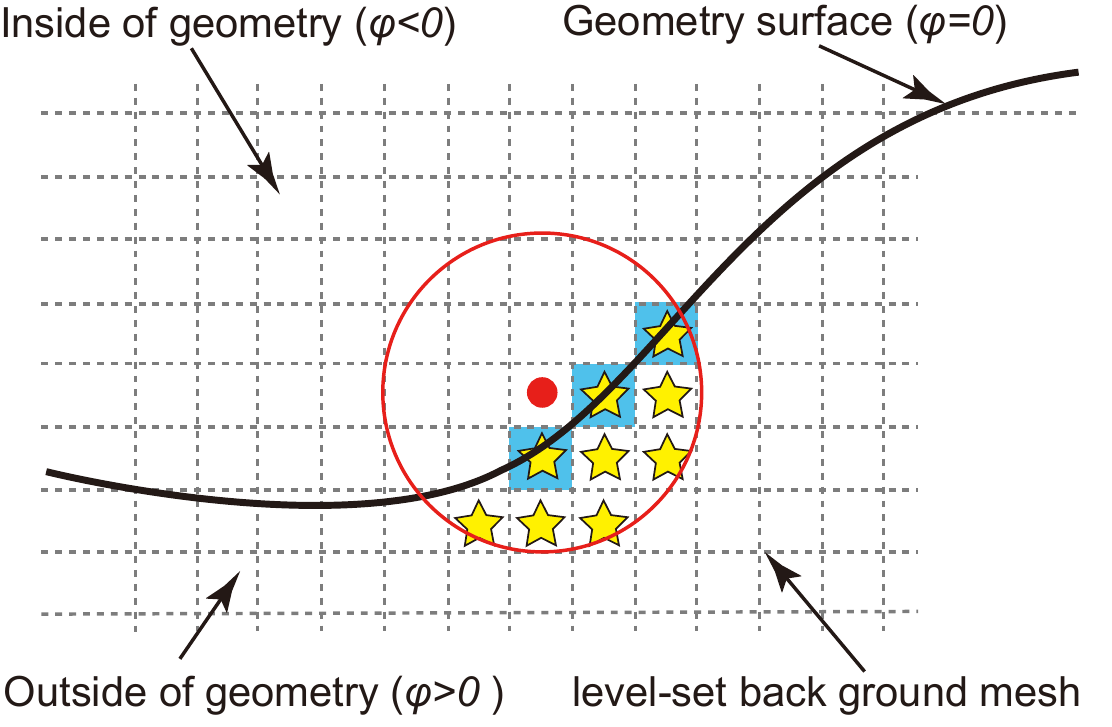}
	\caption{`Static confinement' method for completing kernel of boundary particles. 
		The red circle line is the cut-off region of the red particle $a$ near the geometry surface.
		The white cells inside the geometry and the cut-off region have no contribution 
		to complete the kernel support for the particle $a$.
		The blue cells with a yellow star have the partial volume to complete the kernel support for 
		the particle $a$. The white cells with a yellow star have full volume contributing to the kernel 
		support for the particle $a$. Note that the particle $a$ is just located on the 
		cell-center of the background level-set mesh.}
	\label{figs:Static confinement}
\end{figure}

We first consider an ideal situation in which 
a considered near-surface particle $a$ locates at a cell center, 
as illustrated in Fig. \ref{figs:Static confinement}. 
As its kernel support partially within the surface, 
and partially from the outside region, 
the particle approximation of a derivative $\nabla f_a$ can be obtained by
\begin{equation}\label{eq:discretized particle approximate equations of F}
\nabla f_a\approx \sum_{b, \phi_b < 0} f_{b} \nabla W_{ab} V_b  
+ \sum_{c, \phi_c > 0} f_{c} \nabla W_{ac}V_c , 
\end{equation}
where the second term of the right-hand-side 
provides the support outside of the surface, 
$c$ represent the cell centers of the level set mesh within the cut-off radius from $a$ and $V_c$ are the volume outside of the surface in each mesh cell.
If this approximation is used for particle relaxation as Eq.\ref{eq:momentum},
one can obtain 
\begin{equation}\label{eq:momentum-on-mesh}
\mathbf{F}_{p, a} = -\frac{2 p_{0} V_a }{m_a}\left(\sum_{b, \phi_b < 0} \nabla W_{ab} V_b  
+ \sum_{c, \phi_c > 0} \nabla W_{ac} V_c \right).
\end{equation}
Note that, here, for the volume contribution of each cell near the surface to the extra term introduced in Eq. \ref{eq:momentum-on-mesh}, 
we simply divide the level-set meshes in the support domain into three categories as shown in 
Fig. \ref{figs:Static confinement}. 
The first type is the cells inside both the geometry and cut-off region but not 
cross by geometry surface($\Gamma_0$), which have no contribution to the kernel support. 
The second type is the cells inside the cut-off region and cross by geometry surface($\Gamma_0$), 
which have partial volume (the part outside geometry) to complete the kernel support. 
The last type is the cells that are fully outside geometry but inside the cut-off region,
which have their volume contributed to replenishing the kernel support completely. 
With the level-set method \cite{Method--hu2006,method--sussman1998}, 
the volume fraction for the part outside geometry corresponding to $\phi > 0$ 
of each cell can be estimated by the smoothed Heaviside function 
\begin{equation}\label{eq:heaviside function}
H(\phi,\epsilon)= 
\begin{cases}
0  &\phi < -\epsilon \\
\frac{1}{2} + \frac{\phi}{2\epsilon} + \frac{1}{2\pi}\sin(\frac{\pi\phi}{\epsilon})  &-\epsilon<\phi< \epsilon \\
1  &\phi > \epsilon\\
\end{cases}.
\end{equation}
Then, the extra term in Eq. \ref{eq:heaviside function} can be rewritten as
\begin{equation}\label{eq:kernel gradient from levelset contribution}
	I_{i,j} = \sum_{c, \phi_c > 0} \nabla W_{ac} V_c = \sum_{c, \phi_c > 0} H(\phi_c,\epsilon) l^{m}_f \nabla_a W_{ac},
\end{equation}
here the $l^{m}_f$, $m$ is dimension, 
denotes the volume of each computational cell, $(i,j)$ denotes the cell index.

Note that Eq. \ref{eq:momentum-on-mesh} is only validate when the particles 
$a$ locates on a mesh center. 
For the particle-relaxation process, 
one need the value of the extra term $I_{i,j}$ 
when the particles locates at a general position.
Also note that, this extra term is only dependent on the position 
$(i,j)$ and the surface location, which is fixed during the relaxation process.  
\begin{figure}
	\centering
	\includegraphics[trim = 0cm 0cm 0cm 0cm, clip, width=0.8\textwidth]{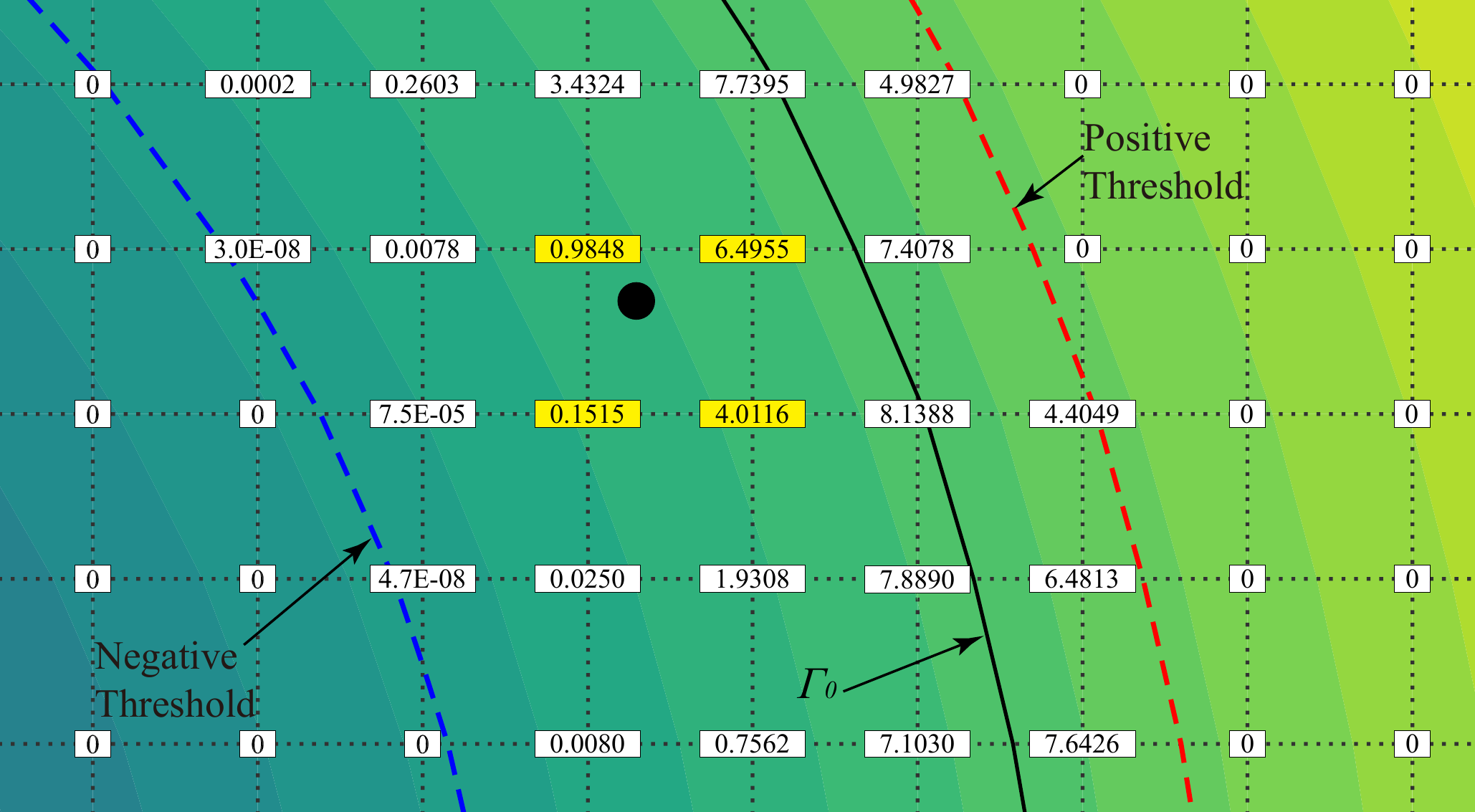}
	\caption{Implementation of the ‘static confinement’ condition. 
		The value of $I_{i,j}$ for the black inner particle can be interpolated  
		from the four values (in yellow boxes) around it. The negative threshold 
		$T_n = -(r_c + l_f)$ and the positive threshold $T_p = l_f$ on both sides of 
		$\Gamma_0$ is set for improving the computational efficiency. Here $r_c$ is 
		the cut-off radius and $l_f$ is the cell-spacing as mentioned in Sec \ref{subsec:levelset-method}.
		Setting $T_p$ as one cell-spacing beyond the geometry is only for the interpolation 
		when the boundary particle is in the cut cell which is crossed by $\Gamma_0$.}
	\label{figs:gradient-kernel-integral}
\end{figure}
Therefore, one can first simply compute the values of $I_{i,j}$ 
at all cell centers near the surface, 
and then obtained the value at an arbitrary particle position, 
with the standard bi- or tri-linear interpolation during the relaxation process, 
(as shown in Fig. \ref{figs:gradient-kernel-integral}). 

By imposing the `static confinement' as the boundary condition for physics-driven 
relaxation process, the distribution of the outer layer particles is significantly 
improved as shown in Fig. \ref{figs:particle_distribution_circle}b. 
By comparing Fig. \ref{figs:particle_distribution_circle}a 
with Fig. \ref{figs:particle_distribution_circle}b, 
it is observed that the outer layer particles can get the same particle spacing 
as the inner particles when the `static confinement' is employed. 
Also, the particles near very sharp features can obtain a better body-fitted distribution as shown in 
Fig. \ref{figs:particle_distribution_airfoil}b. 
The body-fitted particle generation workflow 
with `static confinement' boundary condition is presented in Algorithm \ref{alg:c}. 

\begin{algorithm}[htb!]
    Setup parameters and initialize the physics-driven relaxation\;
    Run Algorithm \ref{alg:a}\;
    \If{Geometry needs to be clean}{
        Run Algorithm \ref{alg:b}\;}
    Generate a preconditioned lattice particle distribution\;
    \While{simulation termination condition is not satisfied}
	{
	 	Get $I_{i,j}$ of each particle by trilinear interpolation\;
	 	Calculate the pressure force $\mathbf{F}_p$ according to 
	 	Eq. \eqref{eq:momentum-on-mesh}\;
	 	Set the time-step $\Delta t$ according to Eq. \eqref{eq:time-step}\;
	 	Update particles position $\mathbf{r}^{n+1}$ according to 
	 	Eq. \eqref{eq:update-position}\;
	 	Get level-set value $\phi_a$ and normal direction $\mathbf N_a$ 
	 	of each particle by trilinear interpolation\;
	 	Constrain particles onto surface according to Eq. \eqref{eq:bounding}\;
		Update the particle-neighbor list and kernel values and gradient \;
		Update the particle configuration \;
	}
	Terminate the simulation.
	\caption{Algorithm for physics-driven relaxation particle generation with `static confinement'}
	\label{alg:c}
\end{algorithm}
%

%
%
\section{Numerical Examples}
\label{sec:numerical examples}
In this section, 
the convergence analysis of average kinetic energy is carried out to verify 
that, without clean-up the non-resolved singular points, the physics-driven 
simulation cannot achieve the steady-state and the convergence. 
In addition, the particle distributions of several complex 3D geometries with 
small structures are exhibited to show the `dirty' geometry clean-up ability 
of the present pre-processing tool, as well as to reveal the importance of 
a cleaned geometry for industry applications. 

In present work, all the simulations below are carried out on an Intel 
Core(TM) CPU i9-9900 3.10GHZ Desktop computer with 64GB RAM and Windows 10 system.   
\subsection{2D airfoil 30P30N}
\label{subsec:airfoil convergence and stability}
As we mentioned in the previous section, the physics-driven process is not 
numerically stable and can not achieve the convergent result 
if there is a non-resolved singular structure present. 
In this section, we consider particle generation for an original 2D airfoil 30P30N 
to qualitatively and quantitatively demonstrate 
the stability and convergence of the present method. 

\begin{figure}
	\centering
	\begin{subfigure}[b]{0.9\textwidth}
		\centering
		\includegraphics[trim = 0cm 0cm 0cm 0cm, clip,width=\textwidth]{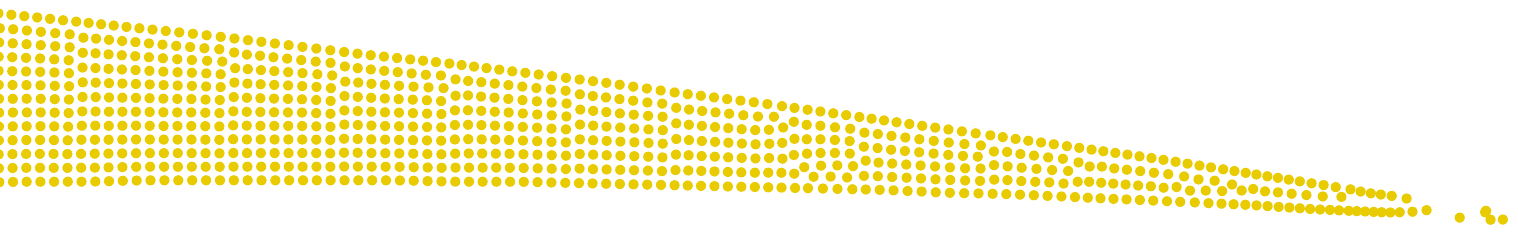}
		\caption{}
		\label{figs:triling edge particles unclean}
	\end{subfigure}
	
	\begin{subfigure}[b]{0.9\textwidth}
		\centering
		\includegraphics[trim = 0cm 0cm 0cm 0cm, clip,width=\textwidth]{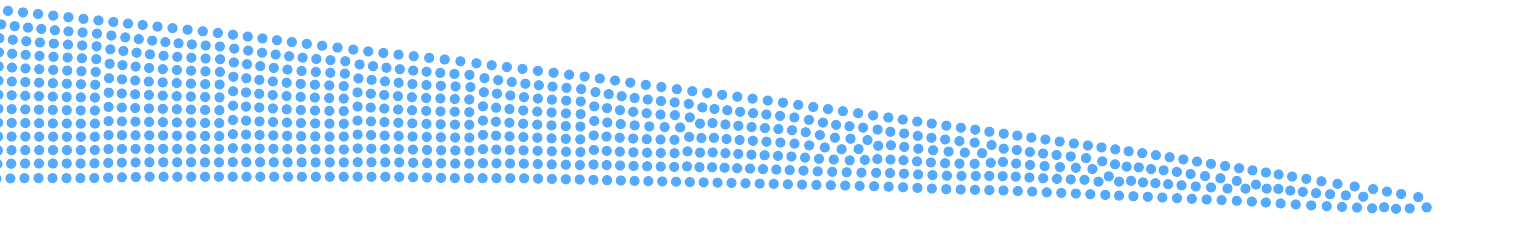}
		\caption{}
		\label{figs:triling edge particles clean}
	\end{subfigure}
	\caption{Particle distribution on trailing edge of the airfoil 
	under the normalized resolution 0.001: 
		(a) Without clean-up non-resolved structure;
		(b) With clean-up non-resolved structure.}
	\label{figs:Particle distribution on trailing with or without clean}
\end{figure}
Fig. \ref{figs:Particle distribution on trailing with or without clean} shows 
the particle distribution at the trailing edge of the airfoil obtained by the 
physics-driven relaxation process with and without clean-up non-resolved structure 
under a normalized resolution of $0.001$. 
As shown in Fig. \ref{figs:triling edge particles unclean}, 
there are still several singular particles at the end of the trailing edge 
that can not reach a stable state during the physics-driven relaxation process, 
free from the main body of the trailing edge and agglomerate together even with a 
relatively high resolution. In contrast, the particles achieve quite good 
body-fitted distribution on the trailing edge after cleaning up the non-resolved 
small fragment as shown in Fig. \ref{figs:triling edge particles clean}. 

\begin{figure}
\centering
    \begin{subfigure}[b]{0.49\textwidth}
         \centering
         \includegraphics[trim = 0cm 0cm 0cm 0cm, clip,width=\textwidth]{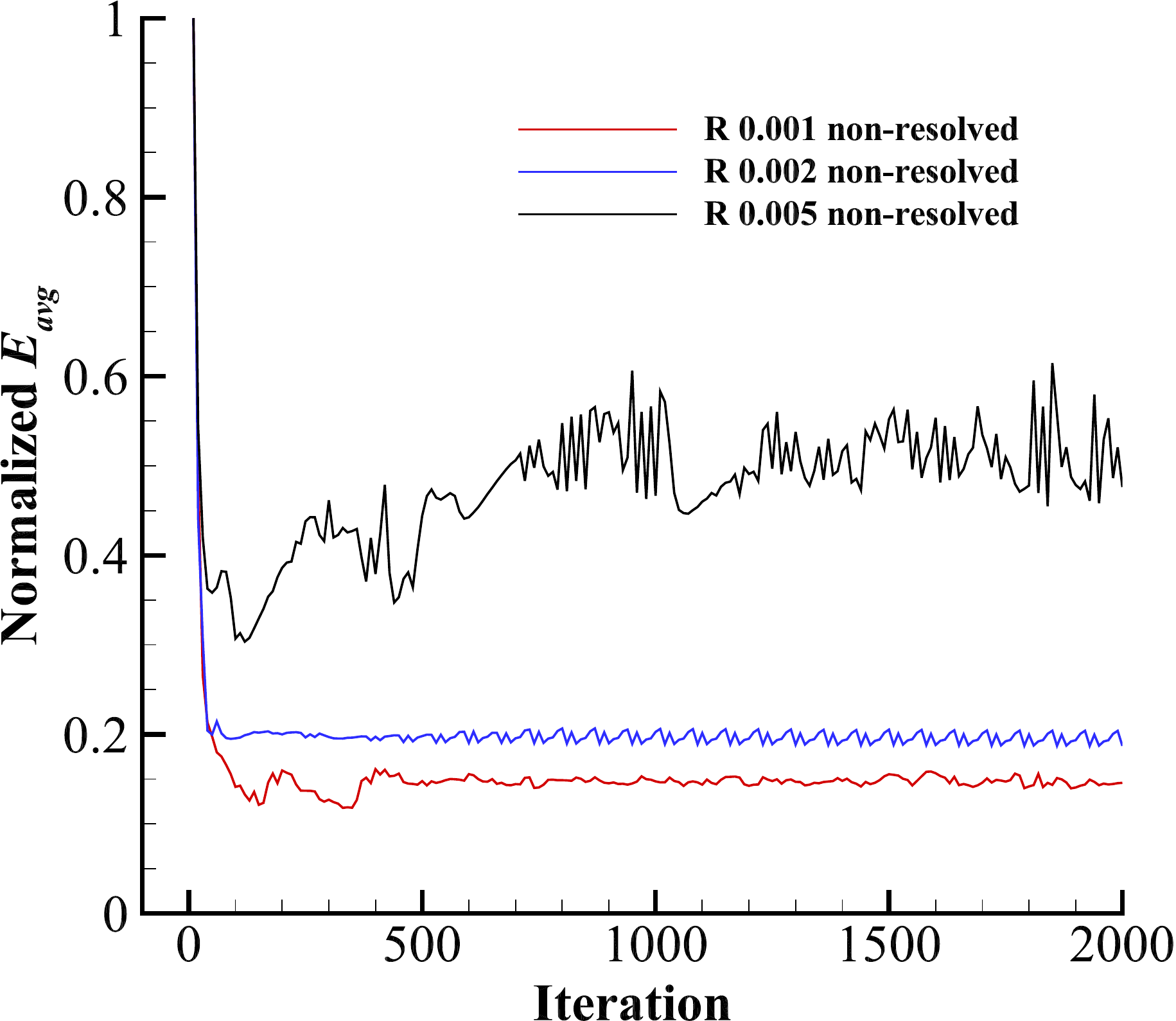}
         \caption{}
		\label{figs:kinetic energy unclean}
     \end{subfigure}
    \begin{subfigure}[b]{0.49\textwidth}
        \centering
         \includegraphics[trim = 0cm 0cm 0cm 0cm, clip,width=\textwidth]{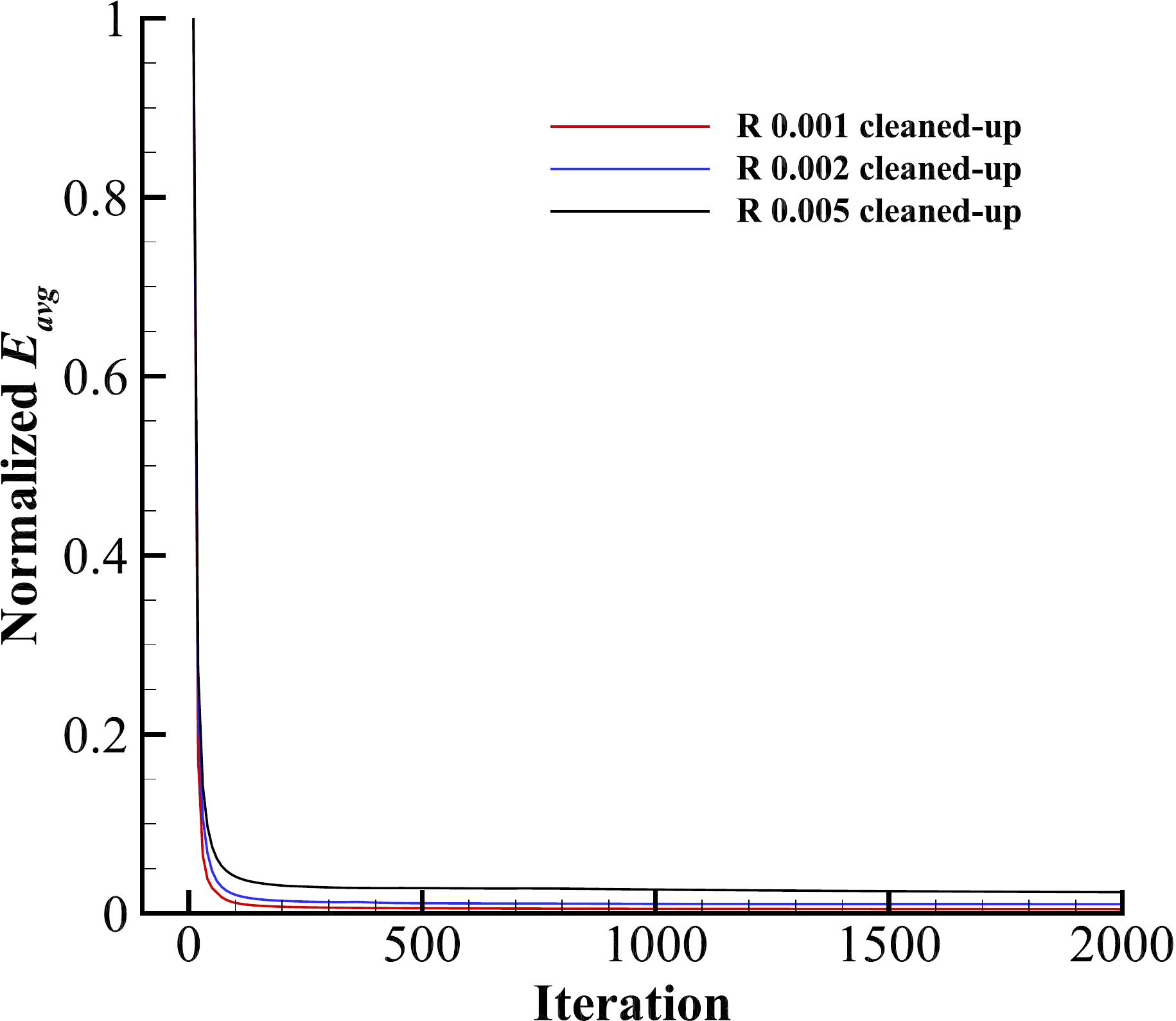}
         \caption{}
		\label{figs:kinetic energy clean}
    \end{subfigure}
 \caption{Average particle kinetic energy during the physics-driven relaxation process: 
 	(a) Without clean-up non-resolved structure;
 	(b) With clean-up non-resolved structure.}
 \label{figs:Average particle kinetic energy}
\end{figure}
Fig. \ref{figs:Average particle kinetic energy} illustrates 
the time history of the average kinetic energy during the physics-driven relaxation 
process with and without clean-up of non-resolved structure. To better explain 
the numerical unstable process, three different normalized resolutions related 
to the airfoil chord are adopted in this test. The three jittery curves in 
Fig. \ref{figs:kinetic energy unclean} show that the physics-driven relaxation 
process is unstable and can not achieve the convergent result as the presence 
of a non-resolved structure. As the resolution decreases, 
the instability becomes more and more severe. In particular,  
the average kinetic energy curve can not even reach a state of periodic 
oscillation when the resolution reduces to 0.005. A coarser grid resolution 
would result in more trailing edge parts with thickness less than a cell-spacing, 
thus generating more mono-layer singular particles. As mentioned before, 
these mono-layer singular particles are the reason for instability and 
even failure to converge. As a comparison in Fig. \ref{figs:kinetic energy clean}, 
after the peak value in the initial status, all the three average kinetic 
energy curves drop rapidly and tend to be stable after 100 iterations with clean-up 
non-resolved structure.
\subsection{SPHinXsys symbol with small structures}
\label{subsec:SPHinXsys symbol}
\begin{figure}
\centering
    \begin{subfigure}[b]{0.49\textwidth}
         \centering
         \includegraphics[trim = 0cm 0cm 0cm 0cm, clip,width=\textwidth]{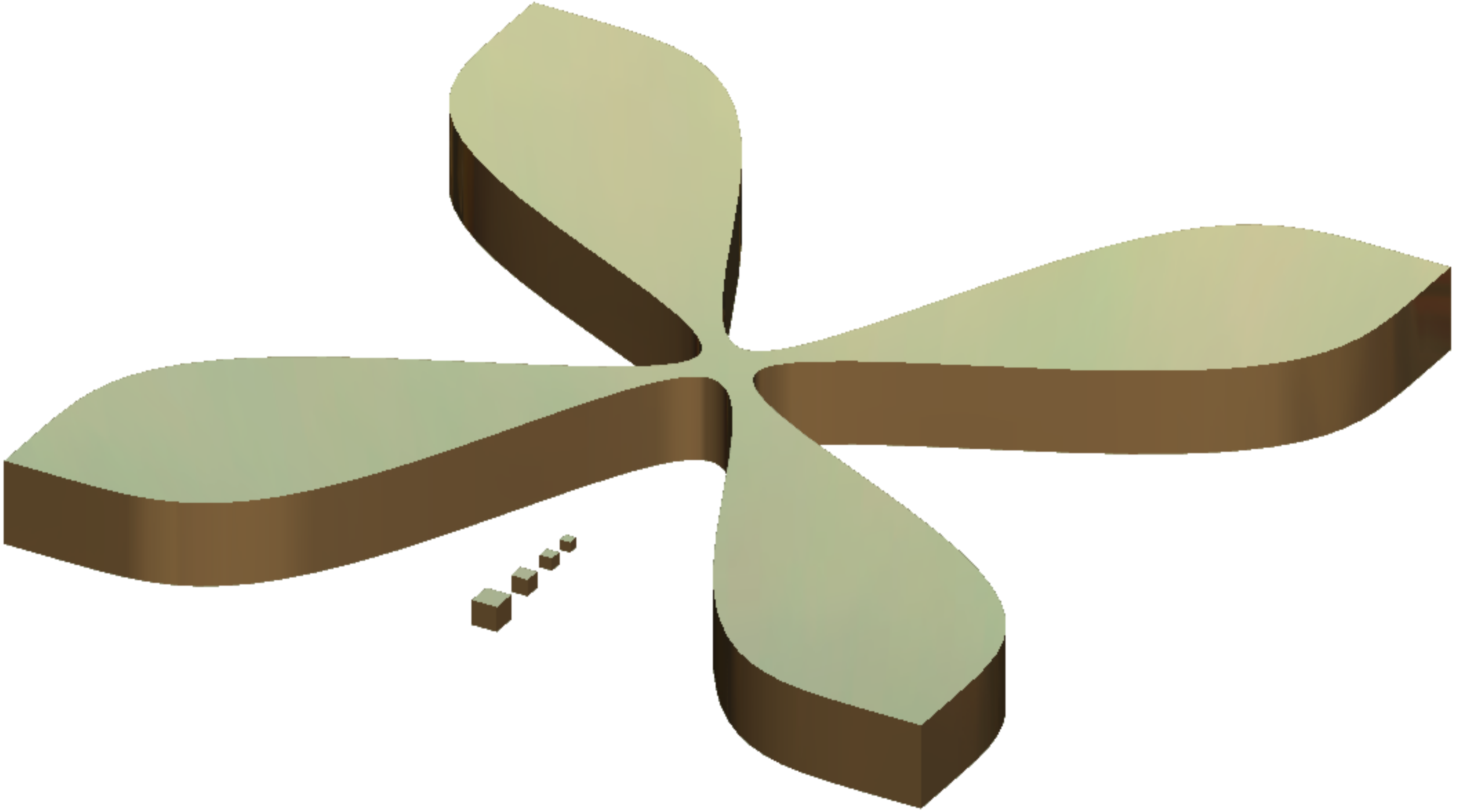}
         \caption{}
		\label{figs:SPHinXsys re-constructed(a)}
     \end{subfigure}
    \begin{subfigure}[b]{0.49\textwidth}
         \centering
         \includegraphics[trim = 0cm 0cm 0cm 0cm, clip,width=\textwidth]{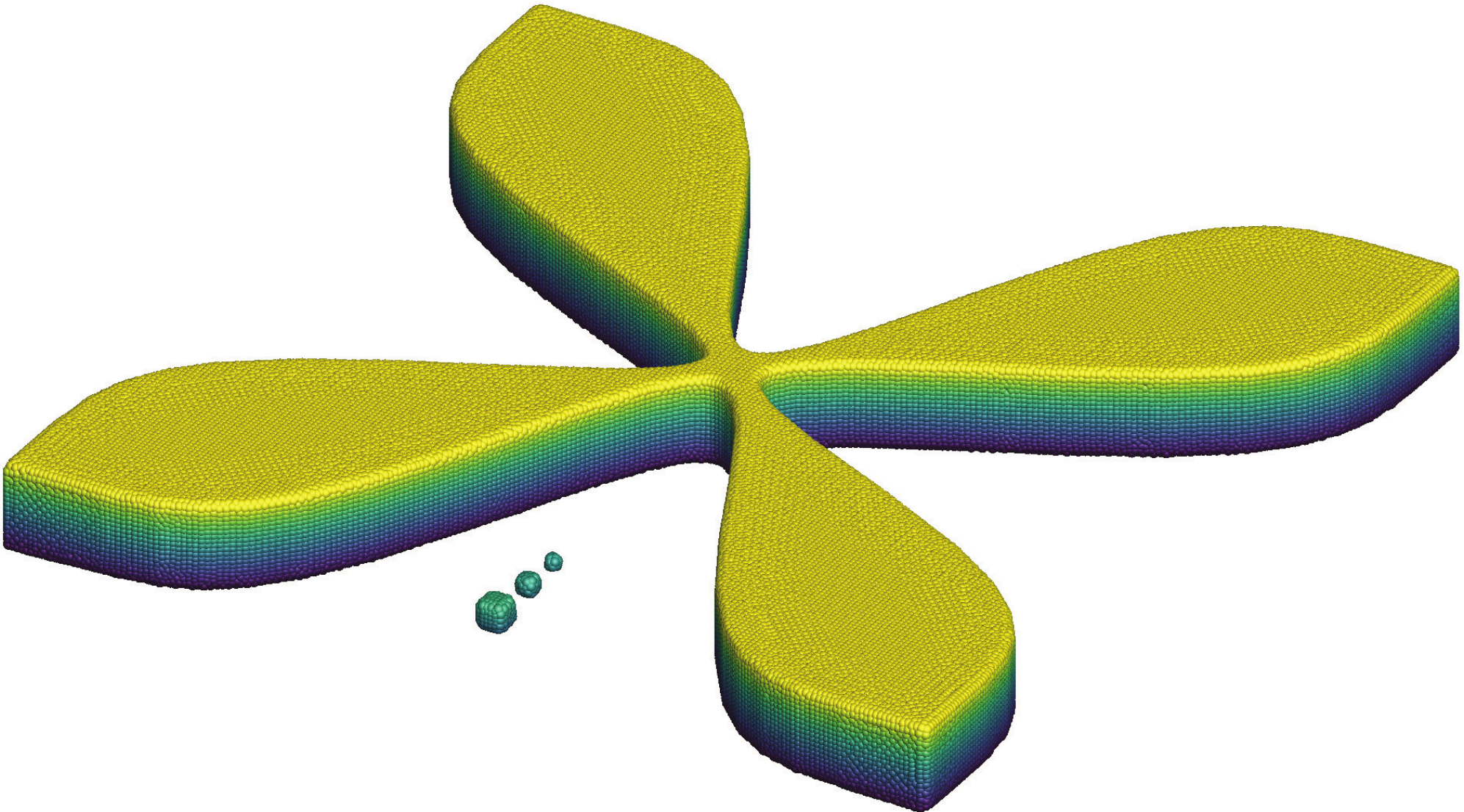}
         \caption{}
		\label{figs:SPHinXsys re-constructed(b)}
     \end{subfigure}
     
    \begin{subfigure}[b]{0.49\textwidth}
        \centering
         \includegraphics[trim = 0cm 0cm 0cm 0cm, clip,width=\textwidth]{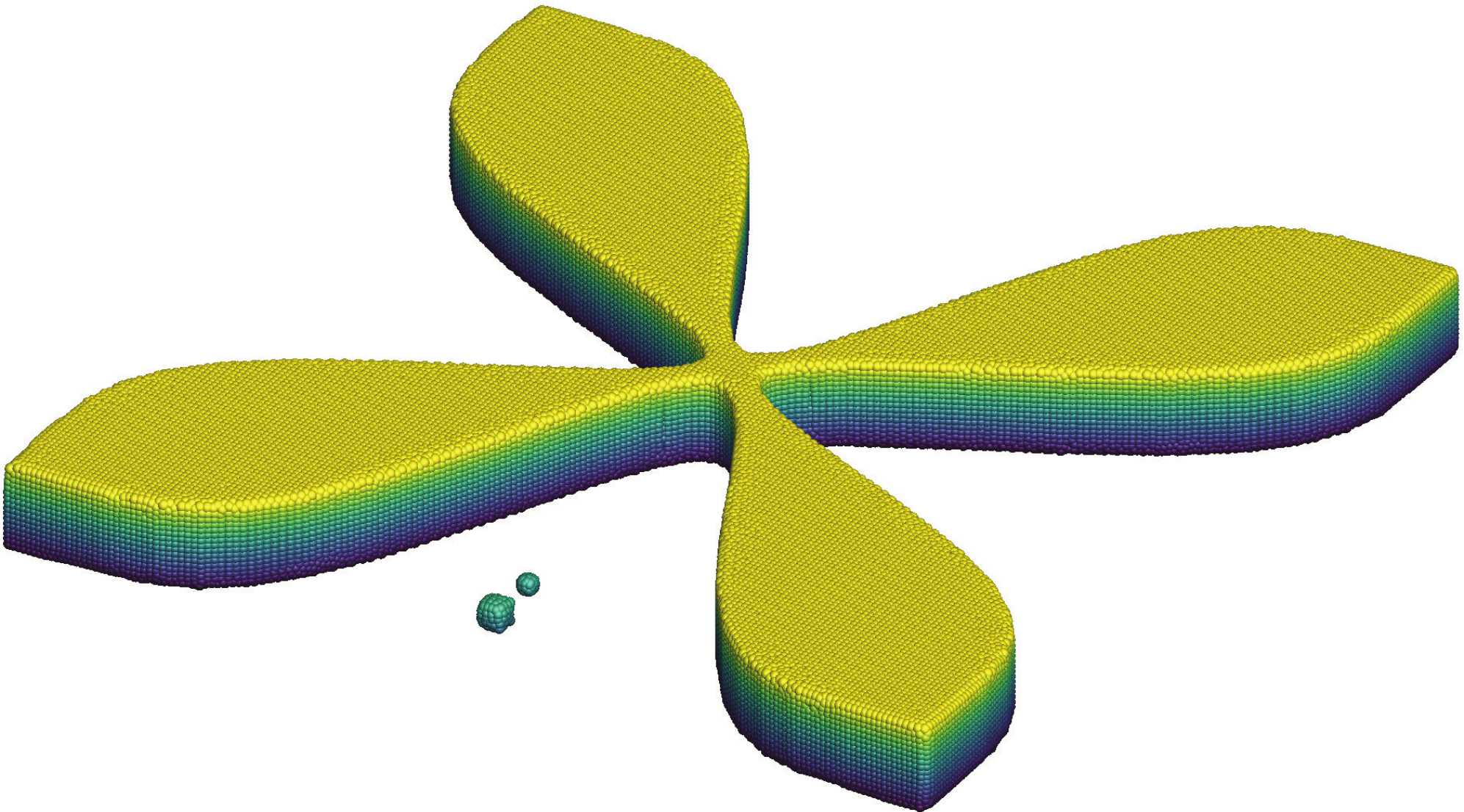}
         \caption{}
		\label{figs:SPHinXsys re-constructed(c)}
    \end{subfigure}
    \begin{subfigure}[b]{0.49\textwidth}
        \centering
         \includegraphics[trim = 0cm 0cm 0cm 0cm, clip,width=\textwidth]{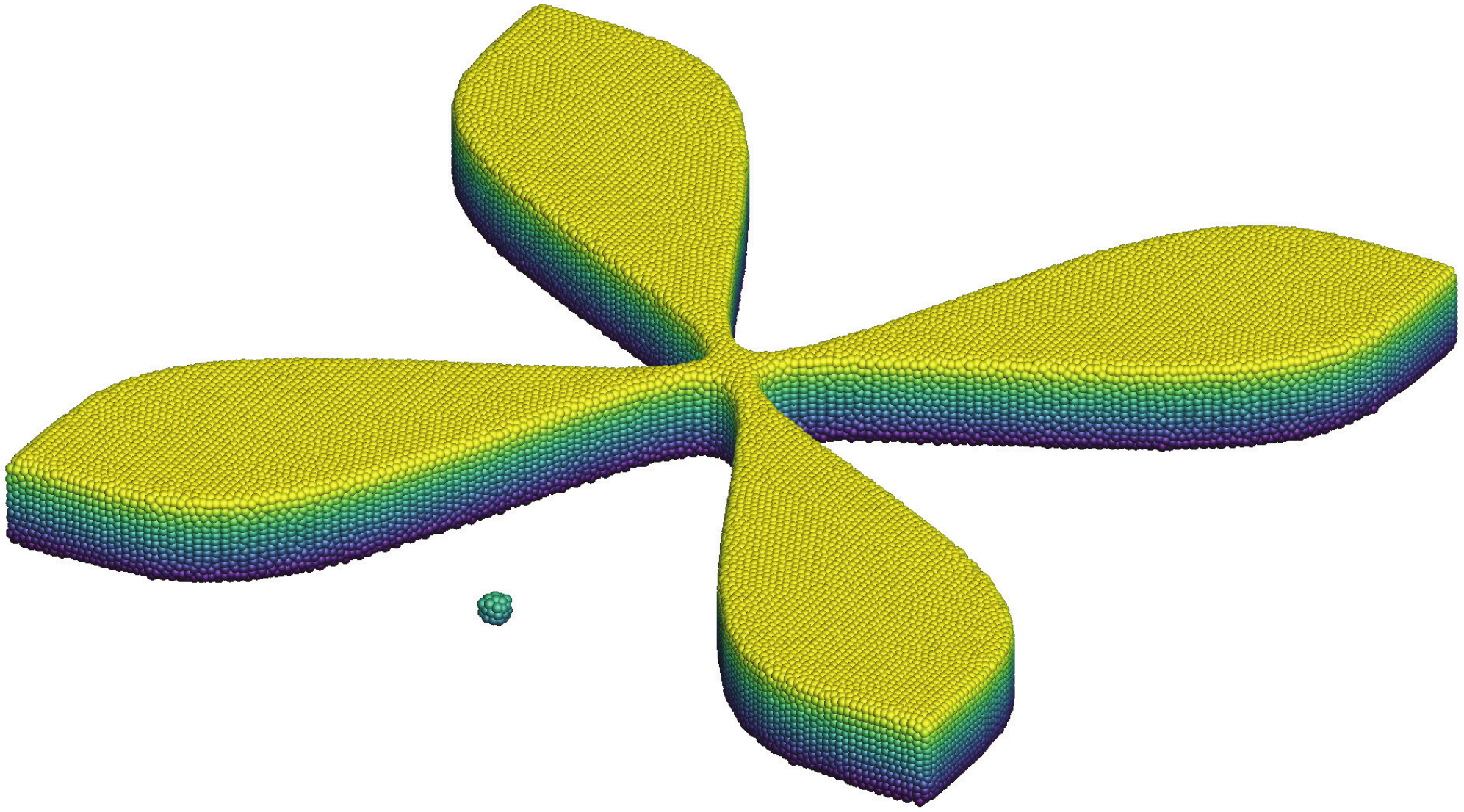}
         \caption{}
		\label{figs:SPHinXsys re-constructed(d)}
    \end{subfigure}
 \caption{Schematic geometry and particles distributions under 
 different spatial resolutions for the 3D SPHinXsys symbol.
 (a) Original geometry surface; (b) Reconstructed with resolution $0.02$;
 (c) Reconstructed with resolution $0.025$; (d) Reconstructed with resolution $0.032$}
 \label{figs:SPHinXsys symbol}
\end{figure}
With the development of 3D scanning technology, 
many geometric models can be obtained through 3D scanning. 
However, 
the scanning process may generate some small structures that are free of 
the main body. These small structures bring challenge for mesh or 
particle generation for computational aspects. In this part, 
we consider generating a particle model for the 3D symbol of our SPHinXsys 
project with some small structures to validate the performance of the 
present pre-processing tool for automatically identifying and cleaning 
up the non-resolved small structures free from the main body 
under diverse spatial resolutions. The schematic geometry of the 
SPHinXsys symbol with small structures are shown in 
Fig. \ref{figs:SPHinXsys re-constructed(a)}.
The size of 3D SPHinXsys symbol is about $4.4 \times 4.4 \times 0.4$, 
together with four small structures of different sizes which are free 
between the two leaves of the main structure.

Fig. \ref{figs:SPHinXsys symbol} also shows the particle distributions 
for the SPHinXsys symbol with different resolutions, 
which are $0.02$, $0.025$ and $0.032$ with respect to $x$ direction.  
It is clear that, with decreasing resolution, 
the free small structures are cleaned-up accordingly. 
Meanwhile, the main body is completely preserved and body-fitted 
particle distributions are obtained.  
In addition, the free small structures that can be resolved at the 
corresponding resolution also produce a body-fitting particle distribution. 
Note that no matter how the resolution is modified, the main body of 
the geometric model is constant, implying the consistent feature of 
a cleaning tool for `dirty' geometry.

\subsection{Skyscraper with a flagpole}
\label{subsec:Building}
In this part, 
a geometric model of a skyscraper with a flagpole is applied 
to test the impact of capturing the small structure on the computational burden. 
The whole computational domain of the skyscraper is about $6 \times 6 \times 21.6$ 
and it includes a two-section flagpole with different diameters. 
The numerical simulations involving this kind of building structure 
are usually aiming at structural strength testing and vibration amplitude detection. 
Thus the flagpoles do not play a decisive influence in the simulation and can be removed. 

\begin{table}[]
\caption{Computational burden comparison among different resolutions of the skyscraper }
\label{table:Computational burden comparison skyscraper}
\begin{tabular}{ccccc}
\hline
\begin{tabular}[c]{@{}c@{}}Normalized\\ Resolution\end{tabular} &
  \begin{tabular}[c]{@{}c@{}}Particles QTY\\ (million)\end{tabular} &
  \begin{tabular}[c]{@{}c@{}}Level-set \\ Time (s)\end{tabular} &
  \begin{tabular}[c]{@{}c@{}}Particle Generation\\ Time (s)\end{tabular} &
  \begin{tabular}[c]{@{}c@{}}Iteration \\ Time (s)\end{tabular} \\ \hline
0.08 & 0.271 & 13.10  & 0.10 & 145.98  \\ \hline
0.05 & 1.559 & 54.08  & 0.53 & 900.38  \\ \hline
0.03 & 4.873 & 145.28 & 1.70 & 2874.13 \\ \hline
\end{tabular}
\end{table}

\begin{figure}
\centering
    \begin{subfigure}[b]{0.24\textwidth}
         \centering
         \includegraphics[trim = 0cm 0cm 0cm 0cm, clip,width=\textwidth]{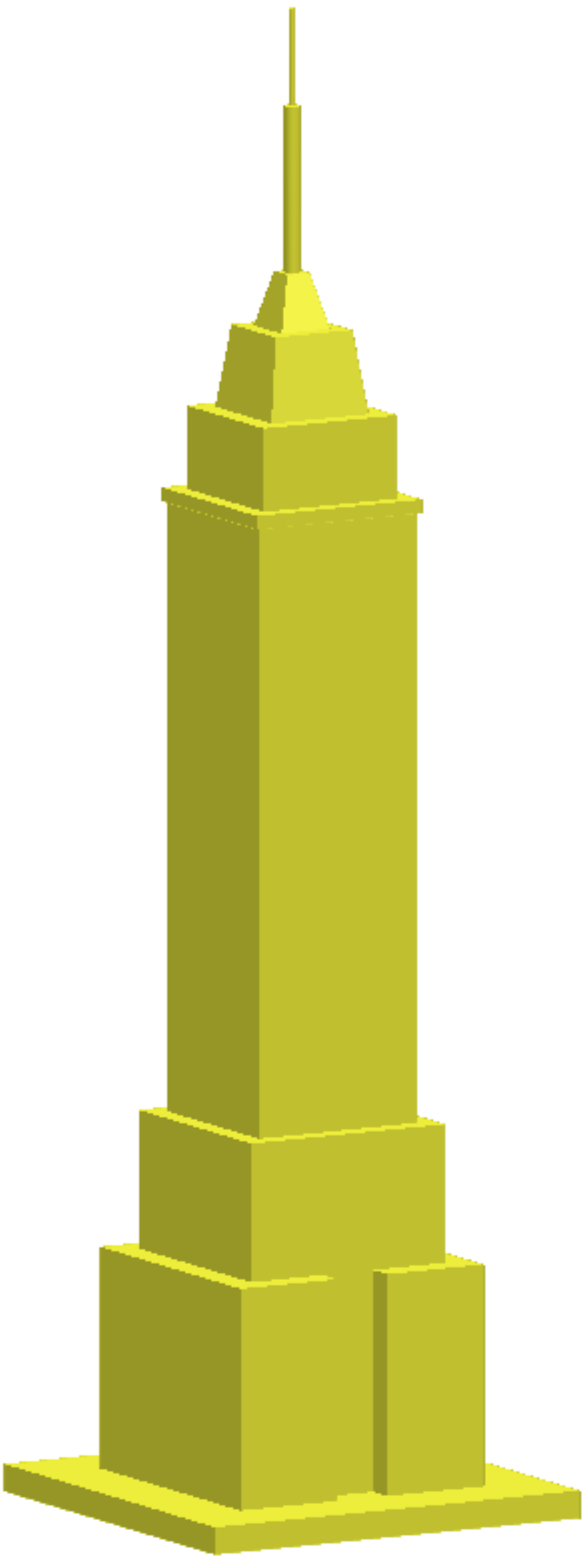}
         \caption{}
		\label{figs:building re-constructed(a)}
     \end{subfigure}
    \begin{subfigure}[b]{0.24\textwidth}
         \centering
         \includegraphics[trim = 0cm 0cm 0cm 0cm, clip,width=\textwidth]{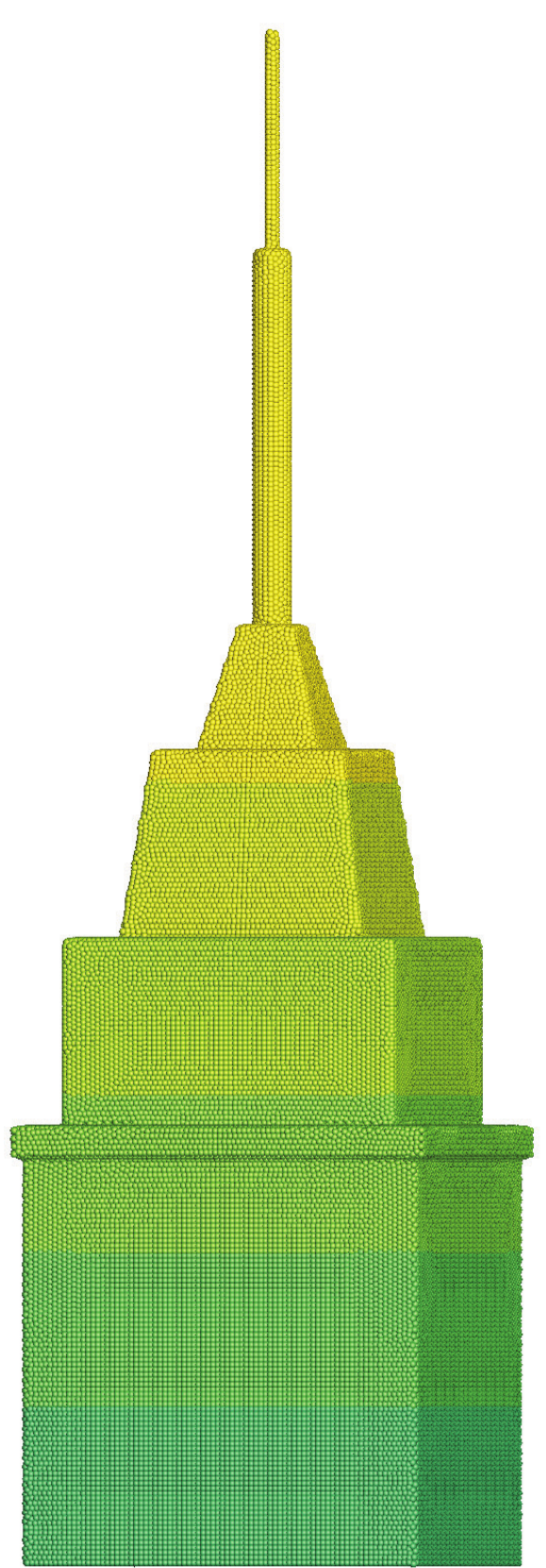}
         \caption{}
		\label{figs:building re-constructed(b)}
     \end{subfigure}
    \begin{subfigure}[b]{0.24\textwidth}
        \centering
         \includegraphics[trim = 0cm 0cm 0cm 0cm, clip,width=\textwidth]{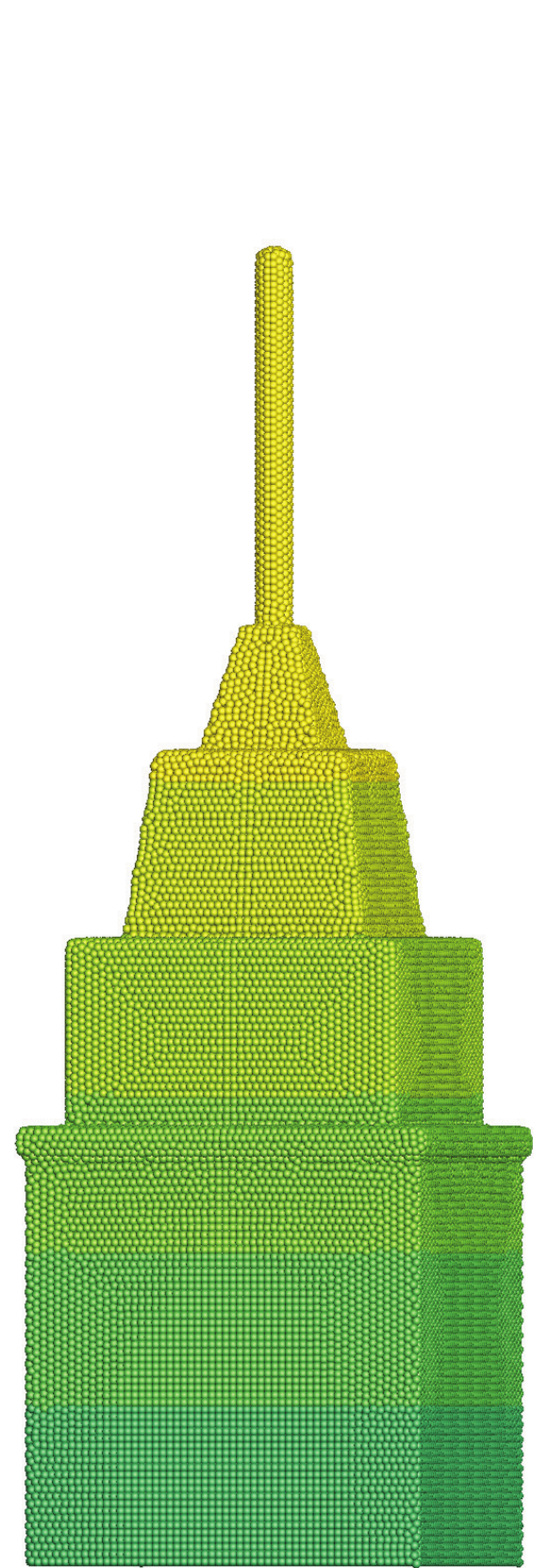}
         \caption{}
		\label{figs:building re-constructed(c)}
    \end{subfigure}
    \begin{subfigure}[b]{0.24\textwidth}
        \centering
         \includegraphics[trim = 0cm 0cm 0cm 0cm, clip,width=\textwidth]{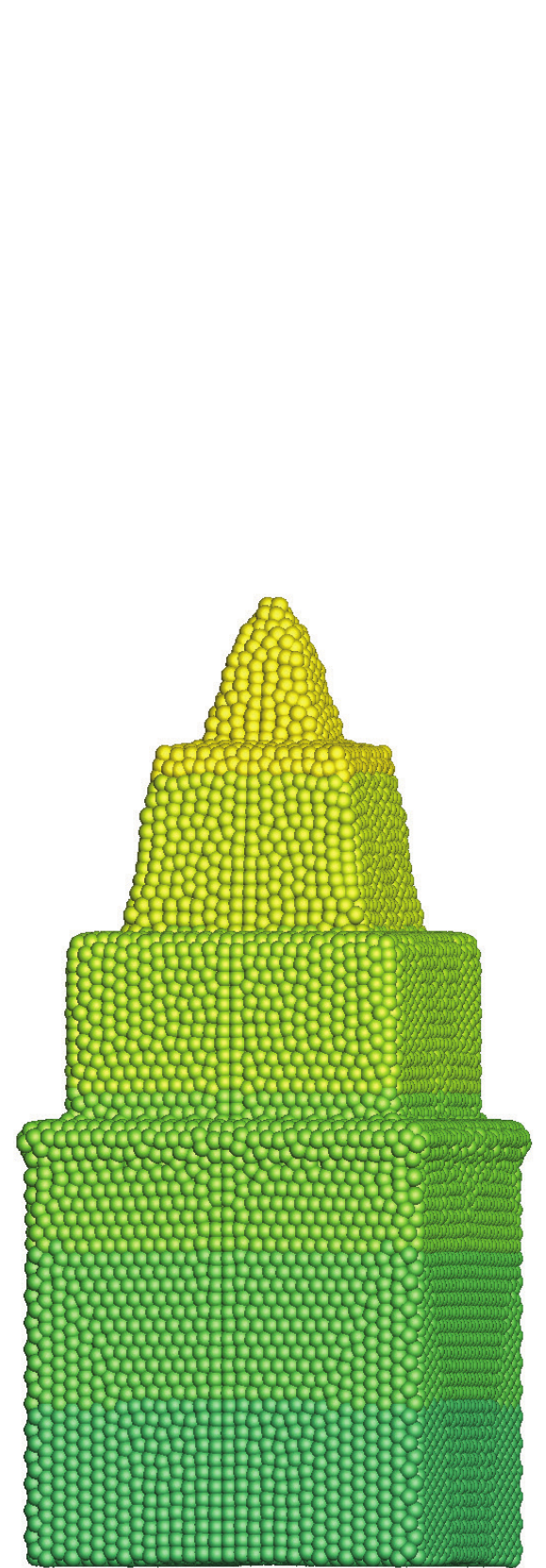}
         \caption{}
		\label{figs:building re-constructed(d)}
    \end{subfigure}
 \caption{The reconstructed geometry surface and particle distribution 
 of a skyscraper under different resolution : 
 	(a) Original geometry model, (b) Reconstructed at resolution of $0.03$, 
 	(c) Reconstructed at resolution of $0.05$ and (d) Reconstructed at resolution of $0.08$. }
 \label{figs:Skyscraper boundary re-constructed}
\end{figure}

Table \ref{table:Computational burden comparison skyscraper} 
presents the computational burden comparison when different resolutions are adopted 
to the numerical simulation of the skyscraper model. 
The numerical process in this example is the physics-driven particle generation process.
The involved level-set operations are level-set initialization, 
`dirty' geometry cleaning-up, and level-set re-initialization.
The `Particle Generation Time' is the 
real-time of lattice distributed particle generation. While the `Iteration Time' is 
the real-time for 1000 iteration steps of the relaxation process. 
The normalized resolutions related to $x$ direction are set as $0.03$, $0.05$ and $0.08$ 
respectively in order to keep the geometric structure in three different states 
(Corresponding to Fig. \ref{figs:building re-constructed(b)} to 
Fig. \ref{figs:building re-constructed(d)}). From the table, 
it is clear that the computational burden and calculation time will explosively 
increase only to capture an insignificant small structure. 
Fig. \ref{figs:Skyscraper boundary re-constructed} further illustrates the 
reconstructed skyscraper features at different resolutions and their particle distributions. 
By omitting different levels of flagpole structure, 
the number of particles in the skyscraper model has been greatly reduced. 
At the same time, its main body is still maintained.

\subsection{Inferior vena cava}
\label{subsec:inferior vena cava}
\begin{figure}
\centering
    \begin{subfigure}[b]{0.85\textwidth}
         \centering
         \includegraphics[trim = 0cm 0cm 0cm 0cm, clip,width=\textwidth]{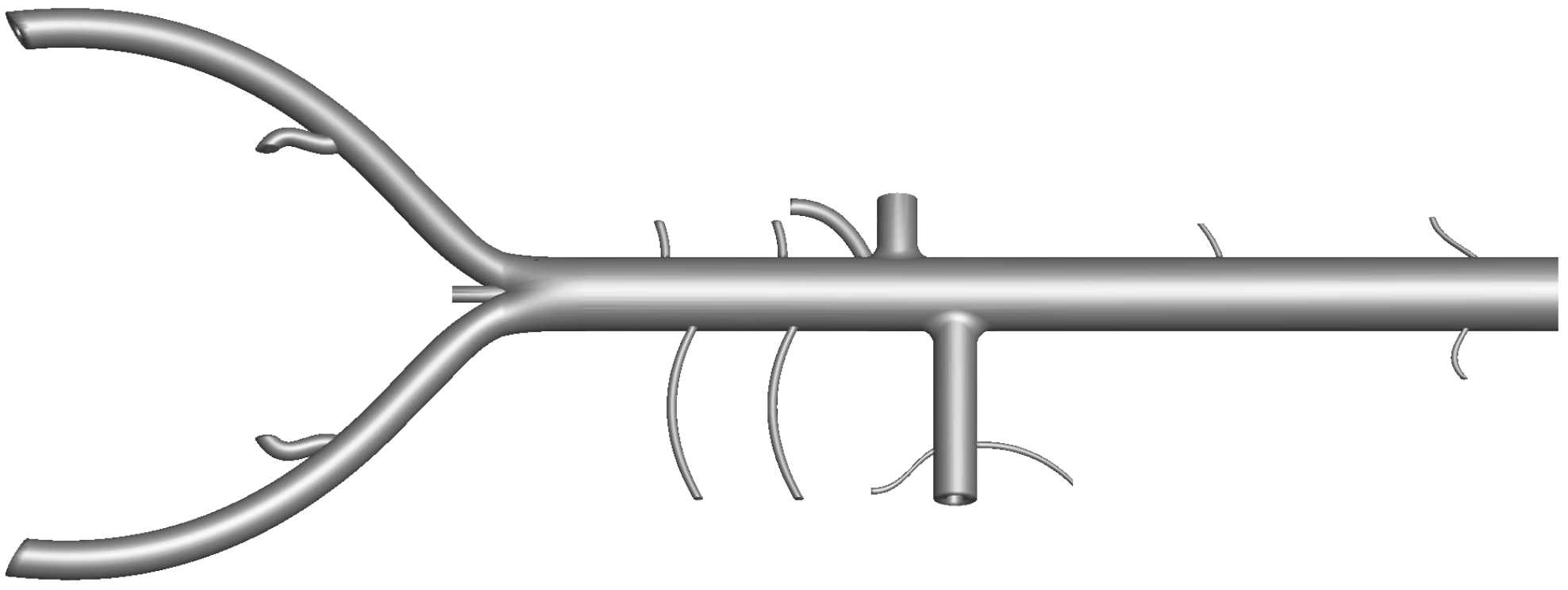}
         \caption{}
		\label{figs:blood vessel re-constructed(a)}
     \end{subfigure}
     
    \begin{subfigure}[b]{0.85\textwidth}
         \centering
         \includegraphics[trim = 0cm 0cm 0cm 0cm, clip,width=\textwidth]{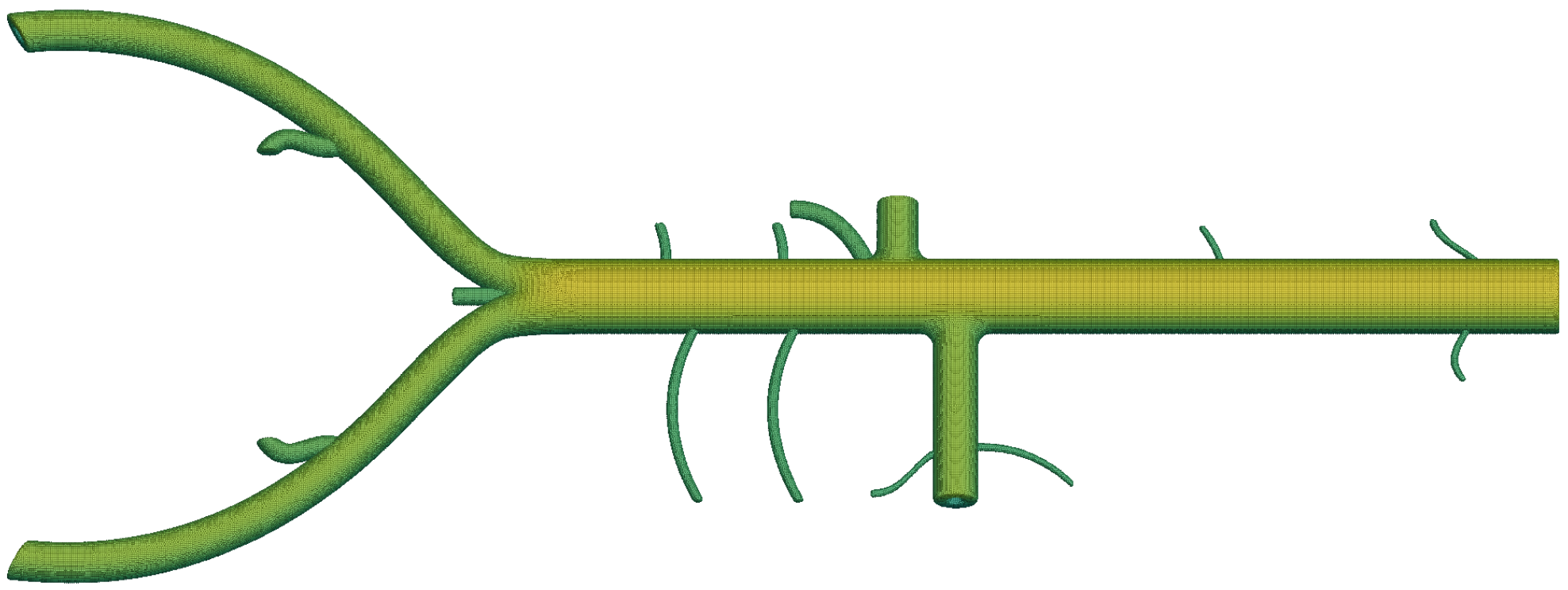}
         \caption{}
		\label{figs:blood vessel re-constructed(b)}
     \end{subfigure}
     
    \begin{subfigure}[b]{0.85\textwidth}
        \centering
         \includegraphics[trim = 0cm 0cm 0cm 0cm, clip,width=\textwidth]{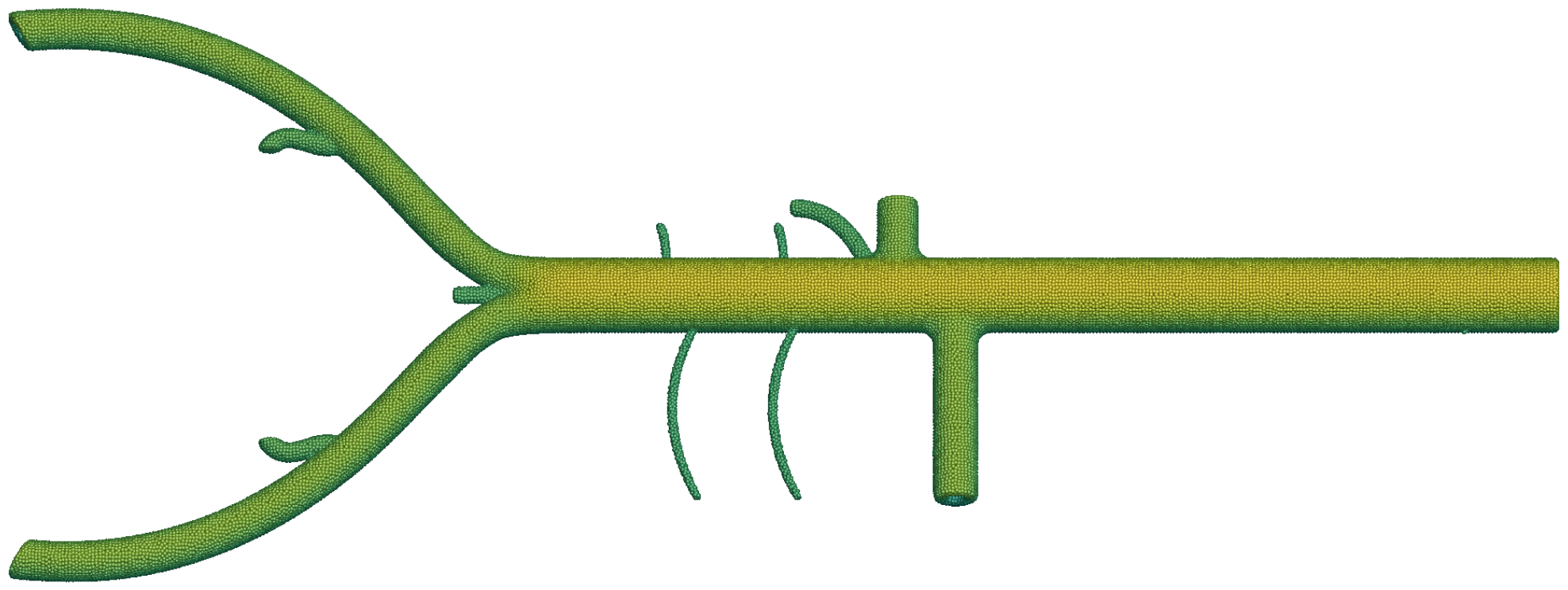}
         \caption{}
		\label{figs:blood vessel re-constructed(c)}
    \end{subfigure}
    
    \begin{subfigure}[b]{0.85\textwidth}
        \centering
         \includegraphics[trim = 0cm 0cm 0cm 0cm, clip,width=\textwidth]{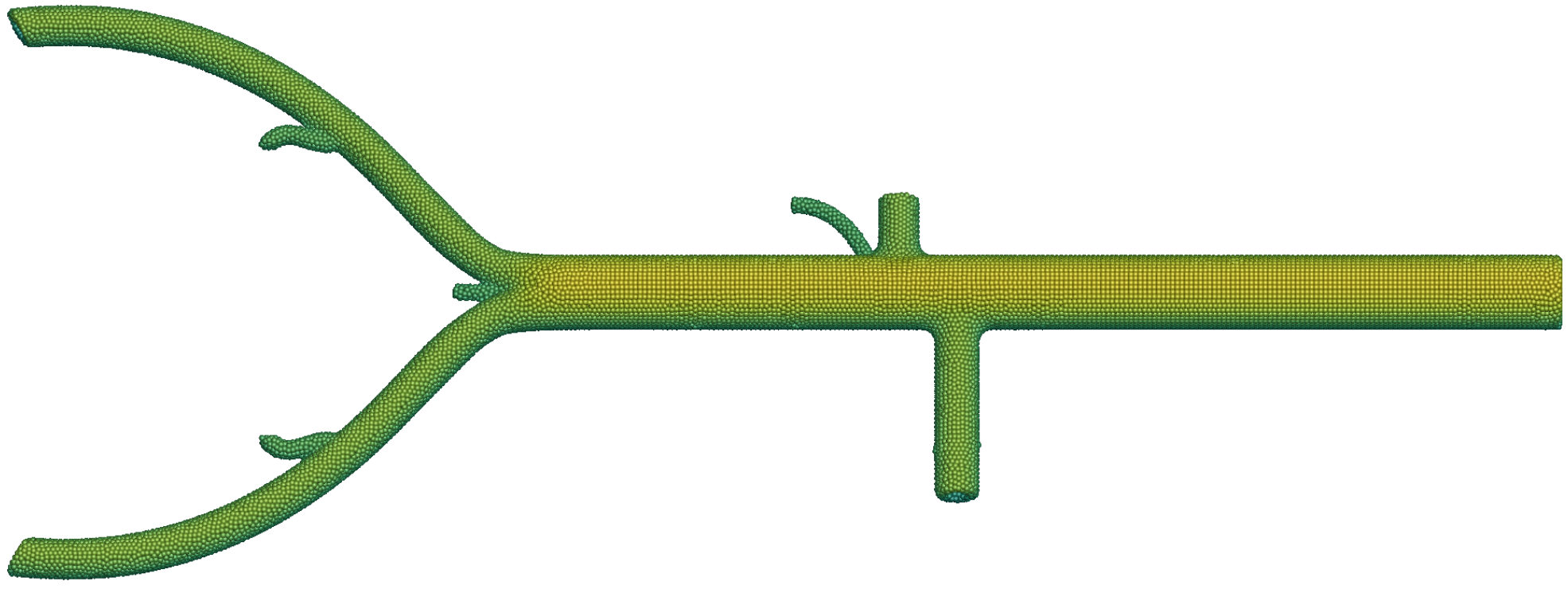}
         \caption{}
		\label{figs:blood vessel re-constructed(d)}
    \end{subfigure}
 \caption{Particle distributions under different resolutions for reconstructed 
 small branches of the inferior vena cava. (a) Original geometry; 
 (b) Reconstructed with resolution $0.55$; (c) Reconstructed with resolution $1.0$;
 (d) Reconstructed with resolution $1.35$.}
 \label{figs:blood vessel geometry boundary re-constructed}
\end{figure}

In this section, 
we consider an inferior vena cava where several branches with different diameters 
are present to test the ability of the proposed method to preserve the small structure 
on the main body under a given resolution. The size of this inferior vena cava 
is around $180 \times 465 \times 33$ and its schematic geometry is shown in 
Fig. \ref{figs:blood vessel re-constructed(a)} 
Generally, this kind of tree structure has many small branches with different diameters. 
For the force study of the main body, these small branches will have no effect 
but increase the computational burden. Therefore, those small branches, 
which cannot be resolved at a given resolution, should be removed. 
While those resolved branches should be kept in the main body at that resolution. 

Fig. \ref{figs:blood vessel geometry boundary re-constructed} 
shows the reconstructed inferior vena cava geometries under different resolutions, 
i.e., $1.35$, $1.0$ and $0.55$ with respect to $x$ direction, 
and the corresponding particle distribution. On the original geometry there 
are several small branches with different diameters attached to the main vessel. 
As the resolution decreases from 0.55 to 1.35 (from 
Fig. \ref{figs:blood vessel re-constructed(b)} to 
Fig. \ref{figs:blood vessel re-constructed(d)}), 
the small branches are gradually cleaned up. However, the main blood vessels 
of the inferior vena cava are still preserved and a body-fitted particle 
distribution is generated.
%
%
\section{Concluding remarks}\label{sec:conclusion}
In this paper, we have developed the level-set based pre-processing techniques
for particle-based methods. Firstly, the `dirty' geometry cleaning technique 
can automatically detect and remove the non-resolved small structures of 
`dirty' geometries at a given grid resolution. A number of numerical tests 
are demonstrated to validate the `dirty' geometry cleaning-up ability of our 
pre-processing tool. The cleaning algorithm of our method can not only identify 
and clean-up the non-resolved small structures but also automatically stop 
cleaning at the main body which has topological consistency. 
In addition, a level-set based `static confinement' 
boundary condition is developed to complete the kernel support in the 
physics-driven relaxation process. By calculating the missing part of kernel 
support and storing it in the background level-set cell center, 
the physics-driven relaxation process obtains the full kernel support 
for particles close to the surface. This allows for better optimization of 
boundary particle distribution, even for those with sharp geometries.         
%
%
\section{Acknowledgement}
Y.C. Yu is fully supported by the China Scholarship Council (CSC) 
(No:201806120023). C. Zhang and X.Y. Hu would like to express 
their gratitude to Deutsche Forschungsgemeinschaft for their sponsorship 
of this research under grant number DFG HU1527/12-4. 
%
%
\section*{CRediT authorship contribution statement}
{\bfseries  Yongchuan Yu:} Investigation, Methodology, Visualization, Validation, 
Formal analysis, Writing - original draft, Writing - review \& editing;
{\bfseries  Yujie Zhu:} Investigation, Methodology, Formal analysis, Writing - review \& editing;
{\bfseries  Chi Zhang:} Investigation, Methodology, Formal analysis, Writing - review \& editing;
{\bfseries  Oskar J. Haidn:} Investigation, Supervision, Writing - review \& editing.
{\bfseries  Xiangyu Hu:} Investigation, Methodology, Supervision, Writing - review \& editing;
%
%
\section*{Declaration of competing interest }
The authors declare that they have no known competing financial interests 
or personal relationships that could have appeared to influence the work reported in this paper.
%
%
\clearpage
\bibliographystyle{elsarticle-num}
\bibliography{levelset_clean}

\begin{thebibliography}{10}
\expandafter\ifx\csname url\endcsname\relax
  \def\url#1{\texttt{#1}}\fi
\expandafter\ifx\csname urlprefix\endcsname\relax\def\urlprefix{URL }\fi
\expandafter\ifx\csname href\endcsname\relax
  \def\href#1#2{#2} \def\path#1{#1}\fi

\bibitem{Industrial--xenakis2017}
A.~Xenakis, S.~Lind, P.~Stansby, B.~D. Rogers, Landslides and tsunamis
  predicted by incompressible smoothed particle hydrodynamics (sph) with
  application to the 1958 lituya bay event and idealized experiment,
  Proceedings of the Royal Society A: Mathematical, Physical and Engineering
  Sciences 473~(2199) (2017) 20160674.

\bibitem{Industrial--cleary2020}
P.~W. Cleary, S.~M. Harrison, M.~D. Sinnott, G.~G. Pereira, M.~Prakash, R.~C.
  Cohen, M.~Rudman, N.~Stokes, Application of sph to single and multiphase
  geophysical, biophysical and industrial fluid flows, International Journal of
  Computational Fluid Dynamics (2020) 1--57.

\bibitem{Industrial--cleary2012}
P.~W. Cleary, M.~Prakash, S.~Mead, X.~Tang, H.~Wang, S.~Ouyang, Dynamic
  simulation of dam-break scenarios for risk analysis and disaster management,
  International Journal of Image and Data Fusion 3~(4) (2012) 333--363.

\bibitem{Industrial--zhang2021}
C.~Zhang, J.~Wang, M.~Rezavand, D.~Wu, X.~Hu, An integrative smoothed particle
  hydrodynamics method for modeling cardiac function, Computer Methods in
  Applied Mechanics and Engineering 381 (2021) 113847.

\bibitem{Industrial--harrison2016}
S.~M. Harrison, R.~C. Cohen, P.~W. Cleary, S.~Barris, G.~Rose, A coupled
  biomechanical-smoothed particle hydrodynamics model for predicting the
  loading on the body during elite platform diving, Applied Mathematical
  Modelling 40~(5-6) (2016) 3812--3831.

\bibitem{Industrial--tanaka2005}
N.~Tanaka, T.~Takano, Microscopic-scale simulation of blood flow using sph
  method, International Journal of Computational Methods 2~(04) (2005)
  555--568.

\bibitem{bio--zhang2022artificial}
C.~Zhang, Y.~Zhu, Y.~Yu, D.~Wu, M.~Rezavand, S.~Shao, X.~Hu, An artificial
  damping method for total lagrangian sph method with application in
  biomechanics, Engineering Analysis with Boundary Elements 143 (2022) 1--13.

\bibitem{Industrial--leroy2014}
A.~Leroy, A new incompressible sph model: towards industrial applications,
  Ph.D. thesis, Universit{\'e} Paris-Est (2014).

\bibitem{Industrial--groenenboom2019}
P.~Groenenboom, B.~Cartwright, D.~McGuckin, O.~Amoignon, M.~Mettichi,
  Y.~Gargouri, A.~Kamoulakos, Numerical studies and industrial applications of
  the hybrid sph-fe method, Computers \& Fluids 184 (2019) 40--63.

\bibitem{Industrial--shadloo2016}
M.~S. Shadloo, G.~Oger, D.~Le~Touz{\'e}, Smoothed particle hydrodynamics method
  for fluid flows, towards industrial applications: Motivations, current state,
  and challenges, Computers \& Fluids 136 (2016) 11--34.

\bibitem{Industrial--lavoie2008}
M.~Lavoie, A.~Gakwaya, M.~N. Ensan, Application of the sph method for
  simulation of aerospace structures under impact loading, in: Proceedings of
  the 10th International LSDYNA Users Conference, Dearborn, Michigan, 2008.

\bibitem{distribute--siemann2019}
M.~Siemann, S.~A. Ritt, Novel particle distributions for sph bird-strike
  simulations, Computer Methods in Applied Mechanics and Engineering 343 (2019)
  746--766.

\bibitem{Industrial--Ortiz2004}
R.~Ortiz, J.~Charles, J.~Sobry, Structural loading of a complete aircraft under
  realistic crash conditions : Generation of a load database for passenger
  safety and innovative design, 2004.

\bibitem{Industrial--zhang2020}
C.~Zhang, Y.~Wei, F.~Dias, X.~Hu, An efficient fully lagrangian solver for
  modeling wave interaction with oscillating wave surge converter, Ocean
  Engineering 236 (2021) 109540.

\bibitem{ocean--luo2021particle}
M.~Luo, A.~Khayyer, P.~Lin, Particle methods in ocean and coastal engineering,
  Applied Ocean Research 114 (2021) 102734.

\bibitem{ocean--cai2022sph}
G.~Cai, M.~Luo, A.~Khayyer, X.~Zhao, Sph simulation of wave impact on coastal
  bridge piers, in: The 32nd International Ocean and Polar Engineering
  Conference, OnePetro, 2022.

\bibitem{Industrial--pan2016}
K.~Pan, R.~IJzermans, B.~Jones, A.~Thyagarajan, B.~van Beest, J.~Williams,
  Application of the sph method to solitary wave impact on an offshore
  platform, Computational Particle Mechanics 3~(2) (2016) 155--166.

\bibitem{Industrial--NASSIRI2016}
A.~Nassiri, B.~Kinsey, Numerical studies on high-velocity impact welding:
  smoothed particle hydrodynamics (sph) and arbitrary lagrangian–eulerian
  (ale), Journal of Manufacturing Processes 24 (2016) 376--381.

\bibitem{Industrial--NDIMANDE2019}
C.~Ndimande, P.~Cleary, A.~Mainza, M.~Sinnott, Using two-way coupled dem-sph to
  model an industrial scale stirred media detritor, Minerals Engineering 137
  (2019) 259--276.

\bibitem{Pre-process-distribute--Dominguez2011}
J.~Dominguez, A.~Crespo, A.~Barreiro, M.~Gesteira, A.~Mayrhofer, Development of
  a new pre-processing tool for sph models with complex geometries, in: {6th
  international SPHERIC workshop}, 2011.

\bibitem{distribute--wang2022centrifugal}
F.~Wang, Z.~Sun, Y.~Sun, K.~Zhang, G.~Xi, Simulation of a centrifugal pump
  based on a lagrangian particle solver, Journal of Fluids Engineering 144~(6),
  061105 (02 2022).

\bibitem{distribute--colagrossi2012}
A.~Colagrossi, B.~Bouscasse, M.~Antuono, S.~Marrone, Particle packing algorithm
  for sph schemes, Computer Physics Communications 183~(8) (2012) 1641--1653.

\bibitem{Industrial--peng2019}
Y.-X. Peng, A.-M. Zhang, F.-R. Ming, S.-P. Wang, A meshfree framework for the
  numerical simulation of elasto-plasticity deformation of ship structure,
  Ocean Engineering 192 (2019) 106507.

\bibitem{FSI--han2018}
L.~Han, X.~Hu, Sph modeling of fluid-structure interaction, Journal of
  Hydrodynamics 30~(1) (2018) 62--69.

\bibitem{FSI--liu2019}
M.~Liu, Z.~Zhang, Smoothed particle hydrodynamics (sph) for modeling
  fluid-structure interactions, Science China Physics, Mechanics \& Astronomy
  62~(8) (2019) 984701.

\bibitem{FSI-MR--zhang2021}
C.~Zhang, M.~Rezavand, X.~Hu, A multi-resolution sph method for fluid-structure
  interactions, Journal of Computational Physics 429 (2021) 110028.

\bibitem{Eulerian--fourtakas2018}
G.~Fourtakas, P.~Stansby, B.~D. Rogers, S.~Lind, An eulerian--lagrangian
  incompressible sph formulation (eli-sph) connected with a sharp interface,
  Computer Methods in Applied Mechanics and Engineering 329 (2018) 532--552.

\bibitem{Eulerian--nasar2019}
A.~Nasar, B.~D. Rogers, A.~Revell, P.~Stansby, S.~Lind, Eulerian weakly
  compressible smoothed particle hydrodynamics (sph) with the immersed boundary
  method for thin slender bodies, Journal of Fluids and Structures 84 (2019)
  263--282.

\bibitem{distribute--2015optimal-initial}
S.~Diehl, G.~Rockefeller, C.~L. Fryer, D.~Riethmiller, T.~S. Statler,
  Generating optimal initial conditions for smoothed particle hydrodynamics
  simulations, Publications of the Astronomical Society of Australia 32 (2015).

\bibitem{distribute--PHL.Groenenboom2014Extended-WVT}
P.~Groenenboom, Particle filling and the importance of the sph inertia tensor,
  in: Proceedings of 9th International SPHERIC Workshop, Paris, France, 2014.

\bibitem{distribute--siemann2014modeling}
M.~Siemann, P.~Groenenboom, Modeling and validation of guided ditching tests
  using a coupled sph-fe approach, in: Proceedings of 9th International SPHERIC
  Workshop, Paris, France, 2014.

\bibitem{MR--fu2019}
L.~Fu, L.~Han, X.~Y. Hu, N.~A. Adams, An isotropic unstructured mesh generation
  method based on a fluid relaxation analogy, Computer Methods in Applied
  Mechanics and Engineering 350 (2019) 396--431.

\bibitem{distribute-MR--ji2020}
Z.~Ji, L.~Fu, X.~Hu, N.~Adams, A consistent parallel isotropic unstructured
  mesh generation method based on multi-phase sph, Computer Methods in Applied
  Mechanics and Engineering 363 (2020) 112881.

\bibitem{MR-distribute--JI2021}
Z.~Ji, L.~Fu, X.~Hu, N.~Adams, A feature-aware sph for isotropic unstructured
  mesh generation, Computer Methods in Applied Mechanics and Engineering 375
  (2021) 113634.

\bibitem{distribute--yujiezhu2021}
Y.~Zhu, C.~Zhang, Y.~Yu, X.~Hu, A cad-compatible body-fitted particle generator
  for arbitrarily complex geometry and its application to wave-structure
  interaction, Journal of Hydrodynamics 33 (2021) 195--206.

\bibitem{mesh-generate--chawner2016}
J.~R. Chawner, J.~Dannenhoffer, N.~J. Taylor, Geometry, mesh generation, and
  the cfd 2030 vision, in: 46th AIAA Fluid Dynamics Conference, 2016, p. 3485.

\bibitem{clean--keiji2021}
K.~Onishi, M.~Tsubokura, Topology-free immersed boundary method for
  incompressible turbulence flows: An aerodynamic simulation for “dirty”
  cad geometry, Computer Methods in Applied Mechanics and Engineering 378
  (2021) 113734.

\bibitem{boundary--2017}
T.~Long, D.~Hu, D.~Wan, C.~Zhuang, G.~Yang, An arbitrary boundary with ghost
  particles incorporated in coupled fem--sph model for fsi problems, Journal of
  Computational Physics 350 (2017) 166--183.

\bibitem{boundary--schechter2012}
H.~Schechter, R.~Bridson, Ghost sph for animating water, ACM Transactions on
  Graphics (TOG) 31~(4) (2012) 1--8.

\bibitem{boundary--vela2019}
L.~V. Vela, J.~M. Reynolds-Barredo, R.~S{\'a}nchez, A positioning algorithm for
  sph ghost particles in smoothly curved geometries, Journal of Computational
  and Applied Mathematics 353 (2019) 140--153.

\bibitem{method--ZHANG/SPHinXsys}
C.~Zhang, M.~Rezavand, Y.~Zhu, Y.~Yu, D.~Wu, W.~Zhang, S.~Zhang, J.~Wang,
  X.~Hu, Sphinxsys: An open-source meshless, multi-resolution and multi-physics
  library, Software Impacts 6 (2020) 100033.

\bibitem{method--zhang2021}
C.~Zhang, M.~Rezavand, Y.~Zhu, Y.~Yu, D.~Wu, W.~Zhang, J.~Wang, X.~Hu,
  Sphinxsys: an open-source multi-physics and multi-resolution library based on
  smoothed particle hydrodynamics, Computer Physics Communications (2021)
  108066.

\bibitem{method--sherman2011}
M.~A. Sherman, A.~Seth, S.~L. Delp, Simbody: multibody dynamics for biomedical
  research, Procedia Iutam 2 (2011) 241--261.

\bibitem{narrow_band--adalsteinsson1995fast}
D.~Adalsteinsson, J.~A. Sethian, A fast level set method for propagating
  interfaces, Journal of computational physics 118~(2) (1995) 269--277.

\bibitem{narrow_band--gomez2005reinitialization}
P.~Gomez, J.~Hernandez, J.~Lopez, On the reinitialization procedure in a
  narrow-band locally refined level set method for interfacial flows,
  International journal for numerical methods in engineering 63~(10) (2005)
  1478--1512.

\bibitem{narrow_band--ye2012multigrid}
J.~Ye, I.~Yanovsky, B.~Dong, R.~Gandlin, A.~Brandt, S.~Osher, Multigrid narrow
  band surface reconstruction via level set functions, in: International
  Symposium on Visual Computing, Springer, 2012, pp. 61--70.

\bibitem{MR-mesh-data--han2014}
L.~Han, X.~Hu, N.~A. Adams, Adaptive multi-resolution method for compressible
  multi-phase flows with sharp interface model and pyramid data structure,
  Journal of Computational Physics 262 (2014) 131--152.

\bibitem{method--litvinov2015}
S.~Litvinov, X.~Hu, N.~A. Adams, Towards consistence and convergence of
  conservative sph approximations, Journal of Computational Physics 301 (2015)
  394--401.

\bibitem{Method--adami2013transport}
S.~Adami, X.~Hu, N.~A. Adams, A transport-velocity formulation for smoothed
  particle hydrodynamics, Journal of Computational Physics 241 (2013) 292--307.

\bibitem{Method--zhang2017generalized}
C.~Zhang, X.~Y. Hu, N.~A. Adams, A generalized transport-velocity formulation
  for smoothed particle hydrodynamics, Journal of Computational Physics 337
  (2017) 216--232.

\bibitem{Method--luo2016}
J.~Luo, X.~Hu, N.~A. Adams, Efficient formulation of scale separation for
  multi-scale modeling of interfacial flows, Journal of Computational Physics
  308 (2016) 411--420.

\bibitem{method--han2015}
L.~Han, X.~Hu, N.~A. Adams, Scale separation for multi-scale modeling of
  free-surface and two-phase flows with the conservative sharp interface
  method, Journal of Computational Physics 280 (2015) 387--403.

\bibitem{method--sussman1998}
M.~Sussman, E.~Fatemi, P.~Smereka, S.~Osher, An improved level set method for
  incompressible two-phase flows, Computers \& Fluids 27~(5-6) (1998) 663--680.

\bibitem{Method--hu2006}
X.~Y. Hu, B.~Khoo, N.~A. Adams, F.~Huang, A conservative interface method for
  compressible flows, Journal of Computational Physics 219~(2) (2006) 553--578.

\end{thebibliography}
%
%
\end{document}